\documentclass[preprint2]{aastex} %a double-column, single-spaced document:

\shorttitle{The \textit{rp}-Process in Neutrino-driven Winds}

\shortauthors{Wanajo}

\begin{document}

\title{The {\boldmath \textit{rp}}-Process in Neutrino-driven Winds}

\author{Shinya Wanajo$^{1,\, 2}$}

\affil{$^1$Research Center for the Early Universe, Graduate School of
   Science, University of Tokyo, Bunkyo-ku, Tokyo 113-0033, Japan}

\affil{$^2$Department of Astronomy, School of Science,
   University of Tokyo, Bunkyo-ku, Tokyo, 113-0033, Japan;
   wanajo@astron.s.u-tokyo.ac.jp}

\begin{abstract}
Recent hydrodynamic simulations of core-collapse supernovae with
accurate neutrino transport suggest that the bulk of the early
neutrino-heated ejecta is proton rich, in which the production of some
interesting proton-rich nuclei is expected. As suggested in recent
nucleosynthesis studies, the rapid proton-capture (\textit{rp})
process takes place in such proton-rich environments by bypassing the
waiting point nuclei with the $\beta^+$-lives of a few minutes via the
faster capture of neutrons continuously supplied from the neutrino
absorption by protons. In this study, the nucleosynthesis calculations
are performed with the wide ranges of the neutrino luminosities and
the electron fractions ($Y_e$), using the semi-analytic models of
proto-neutron star winds. The masses of proto-neutron stars are taken
to be $1.4\, M_\odot$ and $2.0\, M_\odot$, where the latter is
regarded as the test for somewhat high entropy winds (about a factor
of two). For $Y_e > 0.52$, the neutrino-induced \textit{rp}-process
takes place in many wind trajectories, and the \textit{p}-nuclei up to
$A \sim 130$ are synthesized with interesting amounts. However,
$^{92}$Mo is somewhat underproduced compared to those with similar
mass numbers. For $0.46 < Y_e < 0.49$, on the other hand, $^{92}$Mo is
significantly enhanced by the nuclear flows in the vicinity of the
abundant $^{90}$Zr that originates from the $\alpha$-process at higher
temperature. The nucleosynthetic yields are averaged over the ejected
masses of winds, and further the $Y_e$ distribution predicted by the
recent hydrodynamic simulation of a core-collapse
supernova. Comparison of the mass-$Y_e$-averaged yields to the solar
compositions implies that the neutrino-driven winds can be potentially
the origin of light \textit{p}-nuclei up to $A \sim 110$, including
$^{92, 94}$Mo and $^{96, 98}$Ru that cannot be explained by other
astrophysical sites. This parametric study suggests that the
neutrino-induced \textit{rp}-process takes place in \textit{all}
core-collapse supernovae, and likely in collapsar jets or disk winds
formed around a black hole. Although the quantitative prediction of
\textit{rp}-processed abundances requires future detailed
multi-dimensional hydrodynamic simulations with accurate neutrino
transport, the current investigation will serve unique constraints to
the fluid dynamics of the early supernova ejecta.
\end{abstract}


\section{Introduction}

The rapid proton-capture (\textit{rp}) process of nucleosynthesis is
expected to take place in proton-rich compositions with sufficiently
high temperature \citep{Wall81}, which leads to the production of
proton-rich isotopes beyond iron. This is an analogous phenomenon to
the rapid neutron-capture (\textit{r}) process in neutron-rich
compositions, which accounts for the production of about half of the
species heavier than iron \citep[see][for a recent review]{Wana06b}. A
major difference between these two processes is the presence of the
Coulomb barrier for proton capture reactions. This limits the
requisite temperature for this process to a small range, $\sim 1-3
\times 10^9\, \mathrm{K}$, in which the proton capture proceeds during
short time compared to the relevant $\beta^+$-decay lifetimes without
substantial photodisintegrations. To date the thermonuclear explosions
of hydrogen-rich material on accreting neutron stars (X-ray bursts)
have been considered to be the promising astrophysical site associated
to the occurrence of \textit{rp}-process \citep[e.g.,][]{Scha98,
Koik99, Brow02, Woos04}. However, little amount of the nucleosynthetic
product is expected to be ejected, whose contribution to the Galactic
chemical evolution is likely to be negligible.

Recent hydrodynamic studies of core-collapse supernovae that take the
accurate neutrino transport into account have shown that the bulk of
the neutrino-heated ejecta during the early phase ($\lesssim 1\,
\mathrm{s}$) is {\it proton-rich} \citep{Lieb03, Prue05, Bura06,
Kita06, Froh06}. Given this proton richness in the innermost ejecta is
the general characteristic, one might think that core-collapse
supernovae provide suitable physical conditions for the
\textit{rp}-process. If this is true, the study of \textit{rp}-process
in core-collapse supernovae will be of special importance, which no
doubt eject the nucleosynthetic products and thus contribute to the
Galactic chemical evolution. However, there are a number of nuclei
with the $\beta^+$-lives of a few minutes on the \textit{rp}-process
path. These ``waiting point'' nuclei inhibit the production of heavy
proton-rich nuclei beyond the iron-group in core-collapse supernovae
\citep{Prue05}.  The first waiting point nucleus encountered soon
after the abundant $^{56}$Ni is $^{64}$Ge. The half life of its
$\beta^+$-decay is 1.06~min, which is obviously longer than the
dynamic timescale of the innermost ejecta of a core-collapse supernova
($< 1\, \mathrm{s}$).

The situation changes dramatically, however, when neutrino-induced
reactions are included in the nucleosynthesis calculations. In fact,
recent studies by \citet{Froh06, Froh06b}, \citet{Prue06}, and
\citet{Wana06} have shown that the $\beta^+$-waiting points are
bypassed via neutron capture reactions even in proton-rich
environments. This is due to the continuous supply of neutrons from
the anti-electron neutrino absorption by free protons in the early
ejecta that is subject to an intense neutrino flux. As a consequence,
the \textit{rp}-process takes place, which leads to the production of
proton-rich nuclei beyond the iron group. In particular, the
production of some light \textit{p}-nuclei by this neutrino-induced
\textit{rp}-process, or ``\textit{$\nu$p}-process'', is anticipated in
\citet{Froh06b} and \citet{Prue06}. In their works, however, only a
small number of thermodynamic trajectories were considered for the
nucleosynthesis calculations, which were also limited for the very
early ejecta ($\lesssim 1\, \mathrm{s}$ after core bounce).  This is
due to the limitations of the hydrodynamic calculations with accurate
neutrino transport currently available, which led to ``successful''
supernova explosions. It is obvious, however, that more investigations
for various physical conditions are needed for a profound
understanding of the overall picture of this newly discovered
nucleosynthesis process.

In this paper, therefore, the neutrino-induced \textit{rp}-process of
nucleosynthesis in neutrino-driven winds is investigated in some detail,
for various physical conditions obtained from (much simpler)
semi-analytic calculations. A special attention is paid for the
production of light \textit{p}-nuclei including $^{92, 94}$Mo and $^{96,
98}$Ru, which cannot be explained even by the most successful scenario
\citep[i.e., the O/Ne layers in core-collapse
supernovae,][]{Pran90,Raye95}. The production of some light
\textit{p}-nuclei ($^{74}$Se, $^{78}$Kr, $^{84}$Sr, and $^{92}$Mo) in
slightly neutron-rich winds has also been suggested by
\citet{Hoff96}. Hence, the nucleosynthesis calculations are performed
for the wide range of initial electron fractions ($Y_e$) covering
\textit{both} proton-rich and neutron-rich winds. This is of particular
importance to discuss whether the neutrino-driven winds can be the major
production site of these light \textit{p}-nuclei.

The thermodynamic and hydrodynamic trajectories of the neutrino-heated
ejecta are obtained with the semi-analytic models of neutrino-driven
winds (\S~2), which were originally developed for the study of
\textit{r}-process \citep{Wana01}. This study is thus a natural
extension of the previous nucleosynthesis study in \citet{Wana01}, to
take the variation of $Y_e$ into consideration \citep[see
also][]{Wana02, Wana06b}. The masses of proto-neutron stars are taken to
be $1.4\, M_\odot$ and $2.0\, M_\odot$ as in \citet{Wana01}. The latter
is regarded as the test case for (reasonably) high entropy winds (about
a factor of two), which has been proposed as a possible model for the
production of heavy \textit{r}-nuclei \citep{Otsu00, Wana01}. The
nucleosynthesis in each wind trajectory is calculated for the wide range
of $Y_e$ including \textit{both} proton-rich and neutron-rich
compositions (\S~3). The mass-averaged yields over the various neutrino
luminosities for a given initial $Y_e$ are compared with the solar
abundances (\S~4). These results are further averaged over the $Y_e$
distribution of the neutrino-heated ejecta obtained from the recent
two-dimensional hydrodynamic calculation by \citet{Bura06}. These
mass-$Y_e$-averaged yields are compared to the solar compositions to
discuss the contribution of the nucleosynthesis in neutrino-driven winds
to the Galactic chemical evolution. Summary and conclusions of the
current study are presented in \S~5.

\section{Neutrino-driven Wind Models}

After several 100~ms from the core bounce, hot convective bubbles are
evacuated from the surface of a proto-neutron star, and the winds
driven by neutrino heating emerge, as can be seen in some hydrodynamic
simulations of ``successful'' supernova explosions \citep{Woos94,
Bura06}. During this wind phase, a steady flow approximation may be
justified. The wind trajectories in this paper are thus obtained using
the semi-analytic models of neutrino-driven winds in \citet{Otsu00}
and \citet{Wana01, Wana02}, which were developed for the study of
\textit{r}-process \citep[see also][]{Qian96, Card97, Thom01}. Here
the models and some modifications added to them are briefly
described.

In this study, a couple of hydrodynamic studies of ``exploding''
core-collapse supernovae are closely referred, in order to link the
current parametrized models to more realistic ones. One is the
one-dimensional core-collapse simulation of a $20\, M_\odot$
progenitor star in \citet{Woos94}. This is in fact only the case that
followed the wind phase for long duration (up to $\sim 20\,
\mathrm{s}$), which is relevant for the current study. It is
cautioned, however, that the neutrino transport was treated rather
approximately, which makes it difficult to obtain the reliable time
evolution of $Y_e$ in the neutrino-processed material. Note also that
the spatial $Y_e$ distribution cannot be obtained with one-dimensional
simulations. The other is the two dimensional simulation of a
collapsing $15\, M_\odot$ progenitor by \citet{Bura06}, which is also
taken in \citet{Prue05, Prue06}. The simulation was limited to the
first $\sim 0.5\, \mathrm{s}$ after core bounce (and linked to a
one-dimensional simulation up to $\sim 1.3\, \mathrm{s}$), but the
accurate neutrino transport was taken into account. Therefore, the
obtained time variation of $Y_e$ and its spatial distribution may be
reliable.

The system is treated as time stationary and spherically symmetric,
and the radius of the neutron star is assumed to be the same as that
of the neutrino sphere. The physical variables in the wind are then
expressed as functions of the distance $r$ from the center of the
neutron star. The ejected mass by neutrino heating is assumed to be
negligible compared to the mass of the neutron star. Therefore, the
gravitational field in which the neutrino-heated matter moves can be
treated as a fixed-background Schwarzschild metric. The velocity
$v(r)$, temperature $T(r)$, and density $\rho(r)$ can be solved with
their boundary conditions by use of relations based on baryon,
momentum, and energy conservation \citep[eqs.~(1)-(3)
in][]{Wana01}. The source term in the equation of energy conservation
is due to both heating and cooling by neutrino interactions. The
gravitational redshift of the neutrino energies, and the bending of
the neutrino trajectories expected from general relativistic effects,
are explicitly taken into account \citep{Otsu00}. The neutrino
luminosities $L_\nu$ of all neutrino flavors are assumed to be equal,
and the rms average neutrino energies are taken to be 10, 20, and
30~MeV, for electron, anti-electron, and the other flavors of
neutrinos, respectively.

As the boundary conditions at the neutrino sphere, the density is
taken to be $10^{10}\, \mathrm{g\ cm}^{-3}$ and the temperature is
determined so that heating and cooling by neutrino interactions are in
equilibrium.  The mass ejection rate at the neutrino sphere $\dot M$
that determines the initial velocity is set to be $\dot M = 0.99
\times \dot M_c$. Here $\dot M_c$ is the critical value that gives the
transonic solution. The wind with $\dot M_c$ becomes supersonic
through the sonic point, while those with $\dot M < \dot M_c$ are
subsonic throughout. In fact, recent hydrodynamic simulations show
that the fast wind collides with the dense shell of slower ejecta
behind the shock and is decelerated again to be subsonic
\cite[e.g.,][]{Bura06}. On the other hand, the subsonic wind, which
cannot escape by itself from the gravitational potential, must be
enough fast (i.e., $\dot M \approx \dot M_c$) to reach the early
\textit{bubble} ejecta. For example, the wind is too slow to reach the
ejecta behind the shock (at several 1000~km) when $\dot M$ is taken to
be, e.g., $0.9 \times \dot M_c$. Thus, the choice of $\dot M$ above
may be reasonable.

In this study, the neutron star masses $M$ are taken to be $1.4\,
M_\odot$ and $2.0\, M_\odot$. The radius of neutrino sphere is assumed
to be $R_\nu (L_\nu) = (R_{\nu 0} - R_{\nu 1}) (L_\nu/L_{\nu 0}) +
R_{\nu 1}$ as a function of $L_\nu$, where $R_{\nu 0} = 30\,
\mathrm{km}$, $R_{\nu 1} = 10\, \mathrm{km}$, and $L_{\nu 0} = 4
\times 10^{52}\, \mathrm{ergs\ s}^{-1}$, as shown in Figure~1
(\textit{dashed line}). This mimics the early time evolution of
neutrino sphere that can be seen in realistic core-collapse
simulations \citep[][see also Figures~2 and 3]{Woos94,Bura06}. In
Figure~1, $\dot M$ for each $M$ is also shown as a function of $L_\nu$
(\textit{solid line}). The resulted (asymptotic) entropy per baryon
$s$ and the dynamic timescale $\tau_\mathrm{dyn}$ ($\equiv
|\rho/(d\rho/dt)|_{T = 0.5\, \mathrm{MeV}}$) are denoted by thick
solid and dot dashed lines, respectively. As can be seen, the model of
$M = 2.0\, M_\odot$ results in higher $s$, shorter
$\tau_\mathrm{dyn}$, and smaller $\dot M$. In particular, the entropy
is about as twice as that for the $M = 1.4\, M_\odot$ model all the
way, which might lead to the production of the heavy
\textit{r}-process nuclei \citep{Otsu00, Wana01}.

It should be noted that no hydrodynamic studies to date indicate the
formation of such a massive proto-neutron star (with the radius of
$\sim 10\, \mathrm{km}$). On the other hand, a recent hydrodynamic
simulation shows an entropy increase owing to the collision with the
slower preceding ejecta, resulting $s \sim 80\, N_A k$ at early times
\citep[$\sim 1\, \mathrm{s}$,][]{Prue05, Bura06}. Furthermore, some
other mechanisms that increase entropy have been also proposed
\citep[e.g.,][]{Qian96, Thom03, Suzu05, Burr06}. In fact, when it
comes to merely entropy, the model of $M = 2.0\, M_\odot$ is rather
close to the hydrodynamic implication in \citet{Prue05} as far as
during the early epoch. Thus, this model can be regarded simply as a
\textit{reasonably} high entropy case that would be expected in
realistic simulations, rather than an extreme case with a very massive
proto-neutron star. This may be better than to obtain the higher
entropy by multiplying a certain factor to the density or temperature
with no physical basis. It should be stressed that the time variations
of $T$, $\rho$, and $r$ are obtained consistently in the current
study, which are all important for the nucleosynthesis (\S~3.3).

The wind trajectories are calculated for 54 constant $L_\nu$ of
$40-0.5 \times 10^{51}\, \mathrm{ergs\ s}^{-1}$ with the intervals of
$1 \times 10^{51}\, \mathrm{ergs\ s}^{-1}$ and $1 \times 10^{50}\,
\mathrm{ergs\ s}^{-1}$ for $L_\nu \ge 2 \times 10^{51}\, \mathrm{ergs\
s}^{-1}$ and $L_\nu < 2 \times 10^{51}\, \mathrm{ergs\ s}^{-1}$,
respectively. The set of time stationary solutions $v(r)$, $T(r)$, and
$\rho(r)$ are then converted to the time variables $r(t)$, $T(t)$, and
$\rho(t)$ for the nucleosynthesis calculations, where $t$ is the time
that is set to zero at the neutrino sphere for a given wind.

In order to link the series of constant $L_\nu$ trajectories to
realistic time-evolving winds, the post bounce time for each $L_\nu$
is defined as $t_\mathrm{pb}(L_\nu) = t_0\, (L_\nu/L_{\nu 0})^{-1}$,
where $t_0 = 0.2$~s (Figure~1, \textit{dotted line}). This is needed
to determine the ejected mass by each wind (i.e., $\dot M(L_\nu)\,
\Delta t_\mathrm{pb}$) when calculating the mass-averaged
nucleosynthesis yields (\S~4). This gives $L_\nu (t_\mathrm{pb}) =
L_{\nu 0} (t_\mathrm{pb}/t_0)^{-1}$ and $R_\nu (t_\mathrm{pb}) =
(R_{\nu 0} - R_{\nu 1}) (t_\mathrm{pb}/t_0)^{-1} + R_{\nu 1}$ (Figs.~2
and 3), which approximately mimic the hydrodynamic results in
\citet{Woos94}.

The time evolutions of \textit{r}, $T_9$ ($\equiv T/10^9\,
\mathrm{K}$), $\rho$, and $s$ (specific entropy) for selected
(odd-numbered) trajectories are shown in Figures~2 (for $M = 1.4\,
M_\odot$) and 3 (for $M = 2.0\, M_\odot$). In these figures, the time
is set to $t = 0$ at $t_\mathrm{pb} = t_\mathrm{pb}(L_\nu)$ for each
wind. As can be seen in the top panels of these figures, the time
evolutions of radii for $t_\mathrm{pb} > 0.5\, \mathrm{s}$ are very
similar to those found in the hydrodynamic results
\citep{Woos94,Bura06}, at which the current wind solutions may be
justified. On the other hand, the early time evolutions for
$t_\mathrm{pb} < 0.5\, \mathrm{s}$ seem less realistic, in which the
convective bubbles dominate in a realistic simulation
\citep{Bura06}. This may not cause a severe problem, since the
\textit{rp}-process takes place dominantly at $t_\mathrm{pb} > 0.5\,
\mathrm{s}$ as discussed in \S~3.3. Both models ($M = 1.4\, M_\odot$
and $2.0\, M_\odot$) show qualitatively similar thermodynamic
histories as can be seen in Figures~2 and 3. Note, however, that the
entropy is about a factor of two higher for the $M = 2.0\, M_\odot$
model all the way.

\section{Nucleosynthesis in Winds}

\subsection{Reaction Network}

Adopting the wind trajectories discussed in \S~2 for the physical
conditions, the nucleosynthetic yields are obtained by solving an
extensive nuclear reaction network code. The network consists of 6300
species between the proton and neutron drip lines predicted by a recent
fully microscopic mass formula \citep[HFB-9,][]{Gori05}, all the way
from single neutrons and protons up to the $Z = 110$ isotopes. All
relevant reactions, i.e. $(n, \gamma)$, $(p,\gamma)$, $(\alpha,
\gamma)$, $(p, n)$, $(\alpha, n)$, $(\alpha, p)$, and their inverse are
included. The experimental data whenever available and the theoretical
predictions for light nuclei ($Z < 10$) are taken from the REACLIB
compilation (F. -K. Thielemann 1995, private communication). All other
reaction rates are taken from the Hauser-Feshbach rates of BRUSLUB
\citep{Aika05} making use of experimental masses \citep{Audi03} whenever
available or the HFB-9 mass predictions \citep{Gori05} otherwise. The
photodisintegration rates are deduced applying the reciprocity theorem
with the nuclear masses considered. The $\beta$-decay rates are taken
from the gross theory predictions \citep[GT2,][]{Tach90}, obtained with
the HFB-9 $Q_{\beta}$ predictions (T. Tachibana 2005, private
communication). Electron capture reactions on free nucleons as well as
on heavy nuclei are also included \citep{Full82, Lang01}.

Rates for neutrino capture on free nucleons
\begin{eqnarray}
\nu_e + n & \rightarrow & p + e^- \\
\bar \nu_e + p & \rightarrow & n + e^+
\end{eqnarray}
and $^4$He,
\begin{eqnarray}
\nu_e + \mathrm{^4He} & \rightarrow & \mathrm{^3He} + p + e^- \\
\bar \nu_e + \mathrm{^4He} & \rightarrow & \mathrm{^3H} + n + e^+
\end{eqnarray}
as well as for neutrino
spallation of free nucleons from $^4$He,
\begin{eqnarray}
\nu + \mathrm{^4He} & \rightarrow & \mathrm{^3H} + p + \nu' \\
\nu + \mathrm{^4He} & \rightarrow & \mathrm{^3He} + n + \nu'
\end{eqnarray}
are also included \citep{Woos90, McLa96}. Neutrino-induced reactions of
heavy nuclei are not included in this study, which may have only minor
effects \citep{Meye98}. The rates for neutrino reactions above are
expressed as \citep{Qian97}
\begin{eqnarray}
\displaystyle
\lambda_\nu \approx 4.97
\left(\frac{L_\nu}{10^{51}\, \mathrm{ergs\ s}^{-1}}\right)
\left(\frac{\mathrm{MeV}}{\left\langle E_\nu \right\rangle}\right) \nonumber \\
\times\left(\frac{\mathrm{100\ km}}{r}\right)^2
\left(\frac{\left\langle \sigma_\nu \right\rangle}{10^{-41}\, \mathrm{cm}^2}\right)
\quad s^{-1},
\end{eqnarray}
where $\left\langle E_\nu \right\rangle$ and $\left\langle \sigma_\nu
\right\rangle$ are the average energy and cross section of the
neutrino species responsible for the reaction. The time evolution of
$L_\nu$ (that is taken to be constant when deriving the wind
trajectories) is taken into account, according to $L_\nu
(t_\mathrm{pb})$ in \S~2. The distance $r$ from the center of the
neutron star for each wind as a function of time can be seen in
Figures~2 and 3.

\subsection{Initial Compositions and {\boldmath $Y_e$}}

Each calculation is initiated when the temperature decreases to $T_9 =
9$ (where $T_9 \equiv T/10^9\, \mathrm{K}$). At this high temperature,
the compositions in the nuclear statistical equilibrium (NSE) are
obtained (mostly free nucleons and $\alpha$ particles) immediately
after the calculation starts. The initial compositions are thus given
by $X_n = 1 - Y_{ei}$ and $X_p = Y_{ei}$, respectively, where $X_n$
and $X_p$ are the mass fractions of neutrons and protons, and $Y_{ei}$
is the initial electron fraction (number of proton per nucleon) at
$T_9 = 9$. In a core-collapse simulation for long duration
\citep[$\sim 20$~s,][]{Woos94}, it is observed that the initial $Y_e$
takes mostly a constant value ($\approx 0.46$) during the early phase
($t_\mathrm{pb} =$ a few s), and gradually decreases to $\approx
0.38$. This is due to the hardening of the anti-electron neutrino
spectrum relative to that of the electron neutrino at later times
\citep{Qian96}. In order to mimic this effect, $Y_{ei}$ is assumed to
be constant ($Y_{e0}$) for $t_0 < t_\mathrm{pb} \le t_1$ and $Y_{ei}
(t_\mathrm{pb}) = (Y_{e0} - Y_{e1}) (t_\mathrm{pb}/t_1)^{-1} + Y_{e1}$
for $t_\mathrm{pb} > t_1$, where $t_1 = 4$~s and $Y_{e1} = 0.1$. This
gives $Y_{ei}$ for each wind such as $Y_{ei} = Y_{e0}$ and $Y_{ei}
(L_\nu) = (Y_{e0} - Y_{e1}) (L_\nu/L_{\nu 0}) (t_1/t_0) + Y_{e1}$ for
$L_\nu \ge 2 \times 10^{51}\, \mathrm{ergs\ s}^{-1}$ and $L_\nu < 2
\times 10^{51}\, \mathrm{ergs\ s}^{-1}$, respectively.

The core-collapse simulation in \citet{Woos94} shows that the
neutrino-heated ejecta at the first few seconds is \textit{neutron rich}
($Y_e \approx 0.46$). On the other hand, recent hydrodynamic studies
with more accurate neutrino transport show that the bulk of the ejecta
during the early phase is \textit{proton rich} \citep[$Y_e \gtrsim
0.5$,][]{Lieb03, Kita06, Bura06}. The reason of this proton richness is
that, when the degeneracy is lifted, the mass difference between the
proton and neutron favors the proton-rich composition as far as the
differences of luminosities and mean energies between the electron
neutrinos and the anti-electron neutrinos are not significant. As a
result, the electron neutrino capture (eq.~[1]) and the positron capture
(inverse of eq.~[2]) overcome their inverses, resulting in $Y_e > 0.5$
\citep[see][for more detailed discussion]{Hoff96, Froh06, Bura06}. In
this study, the nucleosynthesis calculation for each wind is repeated
for various $Y_{e0}$ ($= 0.45-0.65$) with the interval of 0.005 (41
cases), covering the slightly neutron-rich to very proton-rich
compositions. In Figure~4 (\textit{top panel}), the time variations of
$Y_{ei}$ that determines the initial composition for each wind are shown
for the winds of $M = 2.0\, M_\odot$ with $Y_{e0} = 0.60$ (\textit{thin
solid line}) and $M = 1.4\, M_\odot$ with $Y_{e0} = 0.48$ (\textit{thin
dotted line}), respectively.

Neutrino capture on free nucleons changes $Y_e$ \citep[``$\alpha$
effect'',][]{McLa96, Meye98}. Figure~5 shows $Y_e$ (\textit{thick
lines}) along with the mass fractions (\textit{thin lines}) of neutrons
$X_n$, protons $X_p$, $\alpha$ particles $X_\alpha$, and heavy nuclei
$X_h$ ($Z > 2$) as functions of $T_9$, obtained from nucleosynthesis
calculations for the winds with $M = 2.0\, M_\odot$, $L_\nu = 1 \times
10^{52}\, \mathrm{ergs\ s}^{-1}$, and $Y_{e0} = 0.60$ (W201060,
\textit{solid lines}) and $M = 1.4\, M_\odot$, $L_\nu = 1 \times
10^{52}\, \mathrm{ergs\ s}^{-1}$, and $Y_{e0} = 0.48$ (W141048,
\textit{dotted lines}), respectively\footnote{The wind trajectory with
$M = i.j\, M_\odot$, $L_\nu = kl \times 10^{51}\, \mathrm{ergs\
s}^{-1}$, and $Y_{e0} = 0.mn$ is labeled as W$ijklmn$, hereafter.}. For
the wind W201060, $Y_e$ decreases significantly as the temperature
drops. This is due to the increasing $\alpha$ particles, which continues
as far as neutrons are supplied from the neutrino capture on protons
(eq.~[2]). The electron fraction can be expressed as
\begin{eqnarray}
Y_e \approx X_p + \frac{X_\alpha}{2} + \sum_{Z > 2, A} Z\, Y(Z, A),
\end{eqnarray}
where $Y(Z, A)$ is the abundance (number fraction per nucleon, i.e., the
mass fraction divided by $A$) of the nucleus $(Z, A)$.  For W201060,
alpha particles dominate with a small fraction of protons as soon as the
temperature drops to $T_9 \sim 7$. Hence the second term of equation~(8)
governs $Y_e$, while the third term is negligible. As a result, $Y_e$
decreases toward 0.5. Note that the decrease of $Y_e$ for $T_9 \lesssim
3$ is due to the $(n, p)$ reactions during the \textit{rp}-process phase
and the $\beta^+$-decays after freezeout (\S~3.3). On the other hand,
for W141048, $\alpha$ particles and also heavy nuclei dominate at $T_9
\sim 6$. Thus, $Y_e$ is determined by the second and third terms in
equation~(8). The sum of these two terms are close to the original $Y_e$
value (= 0.48), owing to the slight neutron richness of heavy nuclei. As
a result, $Y_e$ keeps its original value.

This can be seen in Figure~4 (\textit{bottom panel}), in which the
changes of $Y_e$, $\mathit{\Delta} Y_e$, until the temperature drops to
$T_9 = 3$ (at which the \textit{rp}-process begins, see \S~3.3) as
function of $Y_{e0}$, for the winds with $M = 1.4\, M_\odot$
(\textit{dotted line}) and $2.0\, M_\odot$ (\textit{solid line}). The
change is not significant for $Y_{e0} \approx 0.47-0.48$ as described
above. For a larger $Y_{e0}$ case ($> 0.5$), $|\mathit{\Delta} Y_e|$ is
greater for the more massive $M$ ($= 2.0\, M_\odot$) model at later
times ($t_\mathrm{pb} = 4$, i.e., $L_\nu = 2 \times 10^{51}\,
\mathrm{ergs\ s}^{-1}$). This is due to the greater entropy for these
winds (see Figures~1-3) that leave more free protons for $Y_e >
0.5$. For the same reason, $|\mathit{\Delta} Y_e|$ is also greater for a
larger $Y_{e0}$ case. For a smaller $Y_{e0}$ case ($\lesssim 0.47$),
$\mathit{\Delta} Y_e$ takes a positive value owing to the abundant
neutrons. The electron fraction at $T_9 = 3$, defined as $Y_{ef} \equiv
Y_{ei} + \mathit{\Delta} Y_e$, is shown as a function of $t_\mathrm{pb}$
in Figure~4 (\textit{top panel, thick lines}) for the winds of $M =
2.0\, M_\odot$ with $Y_{e0} = 0.60$ (\textit{solid line}) and $M = 1.4\,
M_\odot$ with $Y_{e0} = 0.48$ (\textit{dotted line}). At later times
($t_\mathrm{pb} > 4\, \mathrm{s}$), in which the material is assumed to
be rather neutron rich, $Y_e$ increases significantly owing to the
neutrino effect and takes $\approx 0.34-0.36$ at $t_\mathrm{pb} = 16\,
\mathrm{s}$ ($L_\nu = 5 \times 10^{50}\, \mathrm{ergs\ s}^{-1}$). Note
that the time evolution of $Y_{ef}$ for $M = 1.4\, M_\odot$ with $Y_{e0}
= 0.48$ (Fig.~4, \textit{top panel}, \textit{thick dotted line}) is
similar to the hydrodynamic result in \citet{Woos94}.

\subsection{The {\boldmath \textit{rp}}-Process in Proton-rich Winds}

In the proton-rich compositions with the presence of intense neutrino
flux, the proton capture proceeds beyond the iron group nuclei by
bypassing the $\beta^+$-waiting point nuclei, as shown in recent works
\citep{Froh06b, Froh06, Prue05, Wana06}. This can be clearly seen in
Figure~6, which compares the snapshots of nucleosynthesis at $t =
0.22\, \mathrm{s}$\footnote{In this section, the time is set to $t =
0$ at $T_9 = 9$.}  (when the temperature drops to $T_9 \approx 2$) for
W141060 without (\textit{left panel}) and with (\textit{right panel})
neutrino reactions. Without neutrino reactions, the nuclear flow
cannot advance beyond $^{64}$Ge, whose $\beta^+$ half life is
1.06~minutes\footnote{In the current reaction network, $^{65}$As is
omitted, whose proton separation energy is predicted to be negative in
the HFB-9 mass formula \citep{Gori05}. In principle, the nuclear flow
can proceed through $^{65}$As by two proton capture reaction even with
the (small) negative proton separation energy, owing to the presence
of the Coulomb barrier. Inclusion of this isotope would not change the
current result, however, since the time scale of this two $p$ reaction
is no less than $\sim 1\, \mathrm{s}$ \citep{Brow02}.}. In contrast,
the neutrino capture on abundant protons (eq.~[2]) provides a fraction
of neutrons that immediately suffer $(n, p)$ reactions. Note that the
$(n, p)$ reaction acts as the same as the $\beta^+$-decay of a given
species. As a result, the nuclear flow proceeds to the heavier
proton-rich species (Fig.~6, \textit{right panel}).

It is interesting to note that the proton-to-seed abundance ratio
$Y_p/Y_h$ at this point (Figure~6, \textit{right panel}) is still high
($= 49$) despite the moderate entropy ($55\, N_A k$). This is due to the
slowness of the $\alpha + \alpha + \alpha \rightarrow ^{12}$C reaction
that is the bottleneck for the production of heavy nuclei. This is in
contrast to the neutron-rich compositions, in which the faster $\alpha +
\alpha + n \rightarrow ^9$Be reaction followed by $\alpha$ capture is
efficient. This is why the entropy of a few $100\, N_A k$ is needed for
the \textit{r}-process in neutrino-driven winds \citep{Woos94,
Wana01}. In the proton-rich compositions, however, only a moderate
entropy is sufficient for the occurrence of \textit{rp}-process.

This \textit{neutrino-induced} \textit{rp}-process can be seen more
clearly for the wind models with $M = 2.0\, M_\odot$. This is a
consequence of their higher entropies (see Figures~1-3), which leave
more protons needed for the \textit{rp}-process. Figure~7 shows the
snapshots of nucleosynthesis calculations for the winds W201060
(\textit{top panels}), W200865 (\textit{middle panels}), and W200265
(\textit{bottom panels}) during ($T_9 \approx 2$, \textit{left panels})
and after ($t \approx 1.0$~s, \textit{right panels}) the
\textit{rp}-process phase. For the same $L_\nu$ and $Y_{e0}$ to W141060
(Fig.~6, \textit{right panel}) but with $M = 2.0\, M_\odot$ (W201060),
$Y_p/Y_h$ is three times greater ($= 151$) at $T_9 = 2$ (Fig.~7,
\textit{top left}), owing to its higher entropy ($= 92\, N_Ak$). As a
result, the nuclear flow proceeds to the $Z = 50$ magic nuclei, which
leads to the production of the \textit{p}-nuclei with $A \sim 110$.

The detailed nuclear flow patterns for W201060 when the
\textit{rp}-process is active ($T_9 = 2.5$) are shown in Figure~8. Here,
the nuclear flow for each reaction $i \rightarrow j$ is defined as
\begin{equation}
F_{ij} \equiv \dot Y(i \rightarrow j) - \dot Y(j \rightarrow i)
\quad \mathrm{s}^{-1}.
\end{equation}
In Figure~8, $F_{ij}$ and $Y_i$ (for those greater than $10^{-5}$) are
shown in logarithmic scale by arrows and circles, respectively. As can
be seen, $(p, n)$ reactions play a dominant role to carry away the
nuclear abundances from waiting points, while the contribution of
$\beta^+$-decays (off-centered arrows) is minor.

The time variations of the mass fractions (\textit{thin lines}) of
neutrons $X_n$, protons $X_p$, $\alpha$ particles $X_\alpha$, and heavy
nuclei $X_h$ for W201060 are shown in Figure~9, along with those of the
average mass number of heavy nuclei $A_h$ and the temperature
(\textit{thick lines}), where
\begin{equation}
X_h \equiv \sum_{Z>2, A} X(Z, A)
\end{equation}
and
\begin{equation}
A_h \equiv \frac{1}{X_h} \sum_{Z>2, A} A\, X(Z, A) .
\end{equation}
The results with all neutrino-induced reactions (eqs.~[1]-[6]), neutrino
capture on free nucleons only (eqs.~[1] and [2]), and no
neutrino-induced reactions are denoted by the solid, dashed, and dotted
lines, respectively. The initiation of the \textit{rp}-process can be
clearly seen as the sudden increase of $A_h$ at $T_9 \approx 3$, for the
cases with neutrino reactions (\textit{thick solid} and \textit{thick
dashed lines}). This is due to the presence of neutrons ($X_n \sim
10^{-11}$) by the neutrino capture on abundant protons (eq.~[2]), as can
be seen in Figure~9 (\textit{thin solid} and \textit{thin dashed
lines}).

The neutrino reactions for $^4$He tend to inhibit the
\textit{rp}-process as can be seen in Figure~9, although their effects
are moderate. This is due to the spallation reactions of $^4$He
(eqs.~[5] and [6]), while the capture reactions (eqs.~[3] and [4]) are
not important (because of their smaller cross sections). The resulting
$^3$H emitted from the neutrino spallation of equation~(5) immediately
suffers a $(p, n)$ reaction to $^3$He in the proton-rich
compositions. These $^3$He nuclei, in addition to those from the
spallation reaction of equation~(6), capture abundant $\alpha$ particles
to be $^7$Be, which further capture $\alpha$ particles. As a result, a
fraction of $^4$He assembles into heavier nuclei that act as ``neutron
poison'', resulting in the smaller neutron abundance (Figure~9). This is
a similar mechanism shown by \citet{Meye95}, i.e., the
\textit{r}-process is also inhibited by these neutrino reactions on
$^4$He in the neutron-rich compositions. It should be noted that the
relatively large mean energies of $\mu$ and $\tau$ neutrinos assumed in
this study (according to \citet{Woos94}) may lead to the overestimation
of this effect.

In order to see the gross feature of the neutrino-induced
\textit{rp}-process more clearly, the time variations of the mean
lifetimes for $\beta^+$-decay, $(n, \gamma)$, $(p, \gamma)$, and $(n,
p)$ reactions, and neutrino capture on protons (eq.~[2]) are shown in
Figure~10 (\textit{solid lines}), where
\begin{equation}
\tau_{\beta^+} \equiv
\left[\frac{1}{Y_h}\sum_{Z>2, A} \lambda_{\beta^+} (Z, A)\, Y(Z, A) \right]^{-1},
\end{equation}
\begin{equation}
\tau_{n\gamma} \equiv
\left[\frac{\rho Y_n}{Y_h} \sum_{Z>2, A} \lambda_{n\gamma} (Z, A)\, Y(Z, A) \right]^{-1},
\end{equation}
\begin{equation}
\tau_{p\gamma} \equiv
\left[\frac{\rho Y_p}{Y_h} \sum_{Z>2, A} \lambda_{p\gamma} (Z, A)\, Y(Z, A) \right]^{-1},
\end{equation}
\begin{equation}
\tau_{np} \equiv
\left[\frac{\rho Y_n}{Y_h} \sum_{Z>2, A} \lambda_{np} (Z, A)\, Y(Z, A) \right]^{-1},
\end{equation}
and
\begin{equation}
\tau_{\bar \nu_e p} \equiv {\lambda_{\bar \nu_e p}}^{-1}.
\end{equation}
Here,
\begin{equation}
Y_h \equiv \sum_{Z>2, A} Y(Z, A)
\end{equation}
and $\lambda_{\beta^+}(Z, A)$, $\lambda_{n\gamma}(Z, A)$,
$\lambda_{p\gamma}(Z, A)$, $\lambda_{np}(Z, A)$, and $\lambda_{\bar
\nu_e p}$ are the rates of the corresponding reactions. The rates in
equations~(12)-(15) are averaged over the heavy nuclei, which represent
the lifetimes of the dominant isotopes at a given time. The dotted lines
are the corresponding lifetimes when the sum in equations~(12)-(15) are
taken for $Z \ge 48$ only. With the timescales defined above, the
requisite condition for the neutrino-induced \textit{rp}-process can be
expressed as
\begin{equation}
\tau_{p \gamma} \ll \tau_{np} \ll \tau_{\beta^+}.
\end{equation}
Note that the conditions for the \textit{r}-process and
\textit{classical} \textit{rp}-process are simply $\tau_{n \gamma} \ll
\tau_{\beta^-}$ and $\tau_{p \gamma} \ll \tau_{\beta^+}$, respectively.

Top panel of Figure~10 is the result for the wind trajectory of W201060
that is taken to be the reference case here. As can be seen, the $(n,
p)$ reactions are faster than the $\beta^+$-decays all the way during
the first one second. This shows clearly the importance of the $(n, p)$
reactions for the neutrino-induced \textit{rp}-process. The role of $(n,
\gamma)$ reactions is minor, compared to $(n, p)$ reactions. A
saturation of $A_h$ at $t = \mathrm{a\ few} 100\, \mathrm{ms}$ ($T_9
\sim 2$) in Figure~9 indicates the termination of the neutrino-induced
\textit{rp}-process. At later times, $A_h$ slightly decreases by
photodisintegration.  Note that this termination is \textit{not} due to
the exhaustion of protons, as in the case of \textit{r}-process that
ceases generally by the neutron exhaustion \citep{Wana04}. The
proton-to-seed abundance ratio is still high at later times (e.g.,
$Y_p/Y_h = 77$ at $t =1\, \mathrm{s}$), and the bulk of $(p, \gamma)$
reactions are enough fast throughout (Fig.~10, \textit{solid
line}). However, the $(p, \gamma)$ reactions for the heaviest nuclei ($Z
\ge 48$; Fig.~10, \textit{dotted line}) become slow owing to the
increasing Coulomb barrier, and compete with the $(n, p)$ reactions
(i.e., $\tau_{p \gamma} \sim \tau_{np}$) at $t \sim 0.1\, \mathrm{s}$
($T_9 \sim 2$). This results in the broadening and smoothing of the
abundances for $Z > 40$ in Figure~7 (\textit{top panels}), which is
similar to the ``freezeout effect'' seen in the \textit{r}-process
\citep{Wana04}.

What terminates the \textit{rp}-process is thus the Coulomb barrier of
the heaviest nuclei in the nuclear flow, which breaks the relation
between the first two in equation~(18). The slowness of the $(n, p)$
reactions that compete with the $\beta^+$-decays ($\tau_{np} \sim
\tau_{\beta^+}$) can be another reason to cease the \textit{rp}-process. 
The $(n, p)$ reactions becomes significantly slow owing to the less
active neutrino capture on protons (Figure~10) at later times ($t
\gtrsim 1\, \mathrm{s}$), whose lifetime is proportional to the square
of \textit{r} (eqs.~[7] and [16]). For the current wind trajectory
(W201060), the relation between the latter two in equation~(18) breaks
only at late times ($t \approx 1\, \mathrm{s}$ for $Z \ge 2$ and $t
\approx 7\, \mathrm{s}$ for $Z \ge 48$). This shows that the supply of
neutrons by the neutrino capture on protons is sufficient throughout the
\textit{rp}-process phase.

One may consider that the neutrino-induced \textit{rp}-process
proceeds to much higher atomic number to produce \textit{all} the
\textit{p}-nuclei up to $A \sim 200$, \textit{if} the proton
concentration is much higher owing to, e.g., a larger $Y_{e0}$ or a
higher entropy. This may not be true, however, as can be seen in
Figure~7. For the wind trajectory with a larger entropy ($= 103\, N_A
k$) and $Y_{e0}$ ($= 0.65$, W200865) that results in $Y_{ef} = 0.582$,
$Y_p/Y_h = 427$ at $T_9 = 2$ and the nuclear flow proceeds beyond $Z =
50$ (Figure~7, \textit{middle panels}). In fact, this wind results in
the production of the heaviest \textit{p}-nuclei ($A \sim 130$) in the
current study. However, the flow deviates from the proton-drip line
and reaches the $\beta$-stability line \citep[Figure~7, \textit{middle
right}, see also][]{Prue06}. This is \textit{not} due to the
$\beta^+$-decays of the heaviest nuclei, which are rather slow all the
way ($\tau_{\beta^+} \sim 100\, \mathrm{s}$ for $Z \ge 48$; Fig.~10,
\textit{middle panel}, \textit{dotted line}). This is a consequence of
the slowness of proton capture owing to the higher and higher Coulomb
barrier for the heaviest nuclei in the flow, which competes with both
$(n, p)$ and $(n, \gamma)$ reactions as can be seen in Figure~10
(\textit{middle panel}, \textit{dotted lines}). It should be also
noted that there are $\alpha$ unbound nuclei $^{106-108}$Te ($Z =
52$), which is regarded as the end point of the \textit{classical}
\textit{rp}-process \citep{Scha01}. In addition, the proton capture
also slows substantially when the temperature drops below $T_9 \sim 2$
(Figure~10). Thus, the temperature range in which the neutrino-induced
\textit{rp}-process takes place is strictly limited to $2 \lesssim T_9
\lesssim 3$. This differs significantly from the neutron capture
without Coulomb barrier in case of the \textit{r}-process.

At the later wind phase, the entropy increases significantly, as can
be seen in Figures~1-3. The situation is even worse (but interesting),
however, for the production of proton-rich nuclei. The snapshots of
the nucleosynthesis calculations for the wind trajectory of W200265 is
shown in Figure~7 (\textit{bottom panels}). This wind obtains the
highest entropy ($= 176\, N_A k$) during the proton-rich phase assumed
in this study ($t_\mathrm{pb} \le 4\, \mathrm{s}$, see Fig.~4,
\textit{top panel}). As can be seen, the nuclear flow cannot sustain
its proton richness and runs toward the $\beta$-stability when the
temperature drops to $T_9 \approx 2$ (Figure~7, \textit{bottom left})
despite its extremely high $Y_p/Y_h$ ($= 858$). At later times, the
$(n, \gamma)$ reactions push the flow to the neutron-rich side
(Figure~7, \textit{bottom right}). The reason is that the temperature
falls off to $T_9 \approx 1$ quickly in the later winds (Figure~3). In
addition, high $Y_p/Y_h$ also results in (moderately) high $Y_n/Y_h$
with the presence of intense neutrino flux. As shown in Figure~10
(\textit{bottom panel}), when the temperature is below $T_9 \sim 2$,
the $(p, \gamma)$ reactions for the heaviest nuclei in the flow become
significantly slow, while the $(n, \gamma)$ reactions are still
active. As can be seen in these examples, the neutrino-induced
\textit{rp}-process may not lead to the significant production of
\textit{p}-nuclei beyond $A \sim 130$. It is interesting to note that
many of the nuclei synthesized here (Fig.~7, \textit{bottom right
panel}) are usually assigned as \textit{s}-nuclei or even
\textit{r}-nuclei, which are now synthesized in \textit{proton-rich}
compositions \citep[see also][]{Prue06}.

\subsection{$^{92}$Mo Production in Slightly Neutron-rich Winds}

In slightly neutron-rich compositions ($Y_e \approx 0.47-0.49$), two
dominant quasi-statistical-equilibrium (QSE) clusters are formed around
$A \approx 60$ and $A \approx 90$ during the $\alpha$-process phase
($T_9 \approx 7-4$). At the end of the $\alpha$-process phase, the
matter consists of mostly $\alpha$ particles, the heavy nuclei belong to
these two QSE clusters, and a fraction of free nucleons (Figure~12). As
soon as the $\alpha$-process ceases, the proton capture plays an
important role for the production of the \textit{p}-isotope $^{92}$Mo as
suggested by \citet{Hoff96}, whose origin has been a long-standing
mystery \citep[see][for a recent review]{Arno03}. The nucleosynthesis
calculations covering the slightly neutron-rich winds are thus needed to
estimate their possible contribution to the Galactic chemical evolution
of the light \textit{p}-nuclei.

As shown in Figure~11, a fraction of the most dominant species
$^{90}$Zr ($N = 50, Z = 40$) in the QSE cluster (at $A \sim 90$) flows
into $^{92}$Mo ($N = 50, Z = 42$) through various nuclear reaction
channels. The abundance of $^{92}$Mo reaches $17\%$ of its final value
at this time ($T_9 = 3.5$). Here, the $(p,n)$ reaction plays a
dominant role to carry away the nuclear abundances from $^{90}$Zr. In
Figure~12, the time variations of the mass fractions (\textit{thin
lines}) of neutrons $X_n$, protons $X_p$, $\alpha$ particles
$X_\alpha$, heavy nuclei $X_h$, $^{90}$Zr, and $^{92}$Mo for the wind
trajectory of W141048 are shown, along with those of the average mass
number of heavy nuclei ($A_h$, eq.~[11]) and the temperature
(\textit{thick lines}). Note that the neutrino effect on $Y_e$ is
negligible for this wind as described in \S~3.1, i.e., $Y_{ef} \approx
Y_{e0} = 0.48$. The decrease of $X_p$ slows as the QSE cluster that
contains $^{90}$Zr grows (at $T_9 \approx 5$), in order to balance
with the electron fraction of the cluster ($\approx 0.44$) that is
smaller than that of the wind material ($\approx 0.48$). These protons
are gradually absorbed by the nuclei in this cluster after the
$\alpha$-process phase. As a result, the abundance of $^{92}$Mo
increases, reaching its half final value at $T_9 = 3.2$.

The enhancement of $^{92}$Mo in this way is highly dependent on the
temperature history of the wind after the $\alpha$-process phase. The
production of $^{92}$Mo relies largely upon the slower $(p, n)$
reactions (mean lifetime of about 30~ms at $T_9 = 3.5$) on $^{90}$Zr,
since the photodisintegration impedes the much faster $(p, \gamma)$
reaction at this time (Figure~11). Another channel, that is the $(n,
\gamma)$ reaction on $^{90}$Zr, also contributes but suffers from its
inverse. The $(p, \gamma)$ reaction on $^{90}$Zr becomes effective only
when the temperature drops below $T_9 \approx 2.7$ for
W141048. Therefore, a relatively long cooling timescale of the wind
material is required to obtain a large abundance of $^{92}$Mo
\citep[that is close to its QSE value, see][]{Hoff96}. In the current
study, the subsonic solutions of winds with $\dot M \approx \dot M_c$
are adopted as discussed in \S~2. As a result, the wind is decelerated
when passing over the sonic radius, and thus the decrease of temperature
slows down. For the early winds ($t_\mathrm{pb} < 1\, \mathrm{s}$), this
transition takes place around $T_9 \approx 4-3$. As a result, $^{90}$Zr
suffers $(p, n)$ and $(n, \gamma)$ reactions for longer duration,
resulting in the more production of $^{92}$Mo. Other light
\textit{p}-nuclei, e.g., $^{94}$Mo and $^{96, 98}$Ru, are not
significantly produced in this process, although $^{74}$Se and $^{78}$Kr
are moderately enhanced \citep[see \S~4, see also][]{Hoff96}.

It should be noted that the flow patterns appeared in Figure~11 would
be somewhat different when another nuclear data set were adopted
\citep[see][]{Hoff96}. In fact, most of the relevant reactions here
are taken from the theoretical (Hauser-Feshbach) predictions, although
these nuclear masses are well determined by experiments
\citep{Audi03}. Future determinations of these rates based on
experiments are highly desirable.

Neutrino-induced reactions are not of importance for the production of
$^{92}$Mo in these neutron-rich winds. The results with all
neutrino-induced reactions (eqs.~[1]-[6]), free nucleons only (eqs.~[1]
and [2]), and no neutrino-induced reactions are denoted by the solid,
dashed, and dotted lines, respectively. The dashed and dotted lines
cannot be distinguished, which means that the neutrino capture on free
nucleons plays no role. This is due to their small abundances as can be
seen in Figure~12. The neutrino capture on $^4$He does not play a
significant role, either. The spallation of free nucleons from $^4$He,
on the other hand, slightly enhances the abundance of $^{92}$Mo. The
$^3$H and $^3$He nuclei emitted from the spallation reactions (eqs.~[5]
and [6]) immediately capture abundant $\alpha$ particles to be $^7$Be
and $^7$Li, which further capture $\alpha$ particles. The spalled free
nucleons can survive owing to the faster $\alpha$ capture, resulting in
the slight increase of $^{92}$Mo as can be seen in Figure~12.

\subsection{The {\boldmath \textit{r}}-Process in Neutron-rich Winds}

At later times ($t_\mathrm{pb} > 4\, \mathrm{s}$), the entropy of the
winds increases more than $100\, N_A k$ (Figs.~1-3), and $Y_{ei}$ is
assumed to be neutron rich (Fig.~4, \textit{top panel}). This may
provide the suitable condition for the production of \textit{r}-process
nuclei \citep{Woos94, Wana01}. As discussed in \S~4, the winds with $M =
1.4\, M_\odot$ can be the origin of only the light \textit{r}-process
nuclei (up to $A \sim 130$; Fig.~15, \textit{top panel}). On the other
hand, the heavy \textit{r}-process nuclei (up to $A \sim 190$) can be
produced in the winds with $M = 2.0\, M_\odot$ during the late phase
(Fig.~15, \textit{bottom panel}). This is a consequence of the higher
entropies (Figs.~1-3) and in part the shorter dynamic timescales
(Fig.~1) of the winds, resulting in the higher neutron-to-seed abundance
ratios at the beginning of \textit{r}-processing \citep[$T_9 \sim 3$,
see][for more detail]{Wana01, Wana06b}. Note that the nucleosynthetic
results in the later winds do not contribute to the production of
proton-rich isotopes (\S~4), since $Y_{ei}$ is assumed to quickly drop
to be rather neutron rich (Figure~4, \textit{top panel}).

\section{Contribution to the Galactic Chemical Evolution}

\subsection{Mass-averaged Yields as Functions of {\boldmath $Y_e$}}

In order to estimate the contribution of the neutrino-induced
\textit{rp}-process in neutrino-driven winds to the Galactic chemical
evolution, the nucleosynthetic yields for each $Y_{ei}$ model (41 cases)
are mass-averaged over the 54 wind trajectories weighted by $\dot M
(L_\nu) \mathit{\Delta} t_\mathrm{pb}$. Figure~13 compares the averaged
mass fractions $X_\mathrm{ej}$ of \textit{p}-nuclei with respect to
their solar values \citep{Ande89} as functions of $Y_{ef}$ for $M =
1.4\, M_\odot$ (\textit{top left}) and $2.0\, M_\odot$ (\textit{top
right}) models. Here, $Y_{ef}$, which is approximately the value at the
beginning of the \textit{rp}-process phase (at $T_9 = 3$, \S~3.3), is
taken at $t_\mathrm{pb} = 4\, \mathrm{s}$ (at $L_\nu = 2 \times
10^{51}\, \mathrm{ergs\ s}^{-1}$) as representative of different
$Y_{ef}$. The abundances of $^{64}$Zn and $^{90}$Zr are also plotted for
comparison purposes. The contribution of \textit{p}-nuclei produced
during the later phase ($t > 4\, \mathrm{s}$) is not important, as
discussed in \S~4.2. Figure~13 also shows the results for selected winds
(not mass-averaged) for $L_\nu = 8 \times 10^{51}\, \mathrm{ergs\
s}^{-1}$ ($t_\mathrm{pb} = 1\, \mathrm{s}$, \textit{middle panels}) and
$L_\nu = 2 \times 10^{51}\, \mathrm{ergs\ s}^{-1}$ ($t_\mathrm{pb} = 4\,
\mathrm{s}$, \textit{bottom panels}). The histogram that is taken from
\citet{Bura06} is the asymptotic $Y_e$ distribution $p(Y_e)$ of the
neutrino-processed ejecta during the first 468~ms after core bounce for
a $15\, M_\odot$ progenitor star, obtained by their two-dimensional
hydrodynamic simulation.

As can be seen, a variety of \textit{p}-nuclei are produced with
interesting amounts for $Y_{ef} > 0.5$ models. The heavier
\textit{p}-nuclei (up to $A \sim 130$) appear for the greater $Y_{ef}$
models as well as for the larger $M $ ($= 2.0\, M_\odot$) case (i.e.,
greater entropies, see Figs.~1-3). If we \textit{assume} that the $Y_e$
distribution by \citet{Bura06} holds for the current models, the
neutrino-driven winds may contribute the Galactic production of
\textit{p}-nuclei up to $A \sim 110$ ($^{74}$Se, $^{78}$Kr, $^{84}$Sr, $^{92,
94}$Mo, $^{96, 98}$Ru, $^{102}$Pd, and $^{106, 108}$Cd), which show
considerable enhancements for $Y_{ef} \approx 0.50-0.56$. For the winds
with $L_\nu = 8 \times 10^{51}\, \mathrm{ergs\ s}^{-1}$ (Figure~13,
\textit{middle panels}), it can be seen also that the heavier
\textit{p}-nuclei are highly enhanced as $Y_{ef}$ increases. For $L_\nu
= 2 \times 10^{51}\, \mathrm{ergs\ s}^{-1}$ (Figure~13, \textit{bottom
panels}), however, each \textit{p}-process abundance reaches the maximum
and decreases again as $Y_{ef}$ increases. The reason is that the late
winds in the current study cool down quickly below $T_9 \sim 2$ as can
be seen in Figures~2 and 3. As a result, most of the proton-rich
abundances produced during the \textit{rp}-process phase ($T_9 \approx
3-2$, Fig.~7, \textit{bottom left}) suffer neutron capture to
\textit{neutron-rich} isotopes (Fig.~7, \textit{bottom right}),
regardless of the large $Y_{ef}$ as well as the high entropy.

For $Y_{ef} \approx 0.46-0.49$, some light \textit{p}-nuclei ($^{74}$Se,
$^{78}$Kr, $^{84}$Sr, and $^{92}$Mo) are enhanced with interesting
amounts (Fig.~13). Of particular importance is the production of
$^{92}$Mo as discussed in \S~3.4. \citet{Hoff96} showed that a
neutrino-driven wind having $Y_e \gtrsim 0.484$ cures the $N = 50$
overproduction (mostly $^{90}$Zr), which is replaced with the production
of $^{92}$Mo. This strictly limited $Y_e$ range is due to the
competition between the production of $^{90}$Zr (as the seed of
$^{92}$Mo) in more neutron-rich winds and the concentration of protons
(needed for the $^{92}$Mo production) in less neutron-rich winds. In the
current study, the mass-averaged abundance of $^{92}$Mo with respect to
its solar value overcomes that of $^{90}$Zr for wider range of $Y_{ef}$
($\approx 0.46-0.49$, Fig.~13, \textit{top panels}). The reason is that
the abundances are \textit{mass averaged} over wide range of $L_\nu$,
not from a \textit{single} wind trajectory. In particular, the earlier
winds experience longer durations at $T_9 \approx 4-3$, at which the
$^{92}$Mo abundance is greatly enhanced (\S~3.4). In fact, for a
specific wind, the enhancement of $^{92}$Mo is limited to a smaller
$Y_{ef}$ range as can be seen in Figure~13 (\textit{middle} and
\textit{bottom panels}). In addition, the position of $Y_{ef}$ at which
the $^{92}$Mo abundance takes the maximum shifts toward 0.5 for lower
$L_\nu$ (i.e., later winds). This is a consequence that the later winds
cool down quickly below $T_9 \approx 4-3$, thus the proton richness
(i.e., larger $Y_e$) is more important. In some later winds, relatively
high entropies drive the material to form the QSE cluster that contains
$^{90}$Zr even with small neutron excess ($Y_{ef} \approx 0.49-0.50$),
resulting in the enhancement of $^{92}$Mo (Fig.~13, \textit{middle
right} and \textit{bottom panels}).

The mass fractions of \textit{p}-nuclei with respect to their solar
values \citep{Ande89} are also shown as functions of $L_\nu$ in
Figure~14, for selected winds ($M = 1.4\, M_\odot$; \textit{left
panels}, $M = 2.0\, M_\odot$; \textit{right panels}, $Y_{e0} = 0.48$;
\textit{top panels}, $Y_{e0} = 0.60$; \textit{bottom panels}). For the
winds with $Y_{e0} = 0.48$ ($Y_{ef} \approx 0.48$, as representative of
slightly neutron-rich winds), $^{92}$Mo is greatly enhanced at
relatively early times ($L_\nu > 1 \times 10^{52}\, \mathrm{ergs\
s}^{-1}$, i.e., $t_\mathrm{pb} < 0.8\, \mathrm{s}$). This is due to the
slow decrease of the temperature in the early winds at $T_9 \approx
4-3$, in which $^{92}$Mo is most enhanced (\S~3.4). It should be
cautioned that the steady wind approximation assumed in this study may
not hold during the very early (\textit{bubble}) phase ($t_\mathrm{pb} <
0.5\, \mathrm{s}$). However, it is likely that the material ejected as
\textit{bubbles} expand more slowly \citep[e.g.,][]{Bura06}, and
experiences longer time at $T_9 \approx 4-3$ in which $^{92}$Mo might
enhance to a similar (or more) degree. For the winds with $Y_{e0} =
0.60$ ($Y_{ef} \approx 0.54-0.56$, as representative of proton-rich
winds), \textit{p}-nuclei are enhanced at relatively later times ($L_\nu
< 1 \times 10^{52}\, \mathrm{ergs\ s}^{-1}$, i.e., $t_\mathrm{pb} >
0.8$~s), where the steady wind approximation may hold (Figure~14,
\textit{bottom panels}). This is also consistent to the results by
\citet{Prue06}, which showed no significant production of
\textit{p}-nuclei in the bubbles. Note that no \textit{p}-nuclei are
produced in the very late winds ($L_\nu < 2 \times 10^{51}\,
\mathrm{ergs\ s}^{-1}$, i.e., $t_\mathrm{pb} > 4$~s), in which the
matter is assumed to be rather neutron rich (Fig.~4, \textit{top
panel}).

\subsection{Mass-{\boldmath $Y_e$}-averaged Yields}

In reality, the neutrino-heated matter must have a certain distribution
of $Y_e$ when the multi-dimensional effects are taken into account. The
problem is, of course, the unknown mechanism of core-collapse
supernovae, which governs the hydrodynamic and thermodynamic histories
of the neutrino-driven winds. Among a number of \textit{artificially}
induced explosion models, only the two-dimensional calculation with
accurate neutrino transport by \citet{Bura06} provides us a reliable
$Y_e$ distribution of the neutrino-processed material. The explosion in
their simulation was obtained by omitting the velocity-dependent terms
from the neutrino momentum equation. This led to the increase of the
neutrino energy deposition in the heating region by a few 10\%, which
converted a failed model into an exploding one. The $Y_e$ distribution
$p(Y_e)$ of the neutrino-processed ejecta during the first 468~ms after
core bounce for a $15\, M_\odot$ progenitor star obtained by
\cite{Bura06} is overlaid in Figure~13. The histogram has the maximum at
$Y_e \approx 0.5$ and dominates in the proton-rich side.

To test the contributions of the winds for $M = 1.4\, M_\odot$ and
$2.0\, M_\odot$ models, the mass-averaged yields for each $Y_{e0}$ model
(Fig.~13, \textit{top panels}) are further $Y_e$-averaged ($\sim 2000$
winds in total for each $M$) with $p(Y_e)$, \textit{assuming} this
distribution to be representative of neutrino-driven winds. Needless to
say, there is no guarantee that this distribution holds for the current
wind models (at $t_\mathrm{pb} = 4\, \mathrm{s}$) for both $M = 1.4\,
M_\odot$ and $2.0\, M_\odot$ cases. However, this is only the case
obtained by a multi-dimensional hydrodynamic calculation with
\textit{accurate} neutrino transport that is essential to predict the
$Y_e$ distribution of the neutrino-heated ejecta. It may be, therefore,
interesting to see the possible contributions to the Galactic chemical
evolution with this $p(Y_e)$, keeping in mind that a more consistent
estimation of $p(Y_e)$ will be needed in the future study.

The resulting abundances with respect to their solar values are shown in
Figure~15 for $M = 1.4\, M_\odot$ (\textit{top panel}) and $2.0\,
M_\odot$ (\textit{bottom panel}) models as functions of mass number. The
abundances smaller than $X_\mathrm{ej}/X_\odot < 100$ are omitted
here. The even-$Z$ and odd-$Z$ isotopes for a given element are denoted
by circles and triangles, respectively, which are connected by
lines. The \textit{p}-nuclei are marked with filled symbols. The dotted
horizontal lines indicate a ``normalization band'' \citep{Woos94}
between the largest production factor ($^{100}$Mo for $M = 1.4\,
M_\odot$ and $^{92}$Mo for $M = 2.0\, M_\odot$) and that by a factor of
ten less than that, along with the median value (\textit{dashed
line}). This band is taken to be representative of the uncertainty in
the nuclear data involved.

When compared with the results in which the mass average is only for
$L_\nu \ge 2 \times 10^{51}\, \mathrm{ergs\ s}^{-1}$ (Fig.~17,
\textit{top panels}), it is clear that the later winds for $L_\nu < 2
\times 10^{51}\, \mathrm{ergs\ s}^{-1}$ contribute to only the
\textit{r}-nuclei with $A > 90$ but to \textit{p}-nuclei. Figure~15
implies that there exists a correlation between the production of
\textit{p}-nuclei and \textit{r}-nuclei. That is, the higher entropy
model ($M = 2.0\, M_\odot$) produces heavier \textit{p}-nuclei (up $A
\sim 110$) and \textit{r}-nuclei (up to $A \sim 190$) in a
\textit{single} event. This might be true, since both the
\textit{rp}-process and \textit{r}-process favor the high entropy
conditions. It would be premature to conclude that, however, when
considering the highly uncertain late-time evolution of
neutrino-driven winds. It is interesting to note that the
overproduction of $N = 50$ nuclei $^{88}$Sr, $^{89}$Y, and $^{90}$Zr
seen in the previous \textit{r}-process studies \citep{Woos94, Wana01} are
now replaced with the moderate production of $^{92}$Mo. Figure~16
compares the results when $M$ is replaced to $1.6\, M_\odot$
(\textit{top panel}) or $1.8\, M_\odot$ (\textit{top panel}). The
higher entropy ($\sim 20\%$ for the former and $\sim 50\%$ for the
latter) with respect to the case of $M = 1.4 M_\odot$ (Fig.~15,
\textit{top panel}) results in the production of \textit{p}-nuclei with
higher mass numbers (and of \textit{r}-nuclei). This shows that the moderate
entropy of $\sim 60-80\, N_A k$ in early winds ($\sim 1\, \mathrm{s}$)
is sufficient for the reasonable production of light \textit{p}-nuclei up to
$A \sim 100-110$.

Hereafter, only the winds for $L_\nu \ge 2 \times 10^{51}\,
\mathrm{ergs\ s}^{-1}$ ($t_\mathrm{pb} \le 4\, \mathrm{s}$) are
considered, in which the \textit{p}-nuclei are produced. Note that the
production of $^{64}$Zn that is the dominant stable zinc isotope
\citep[but its origin is unknown, see][]{Hoff96, Umed02, Prue05,
Froh06} is about 10 times smaller than the lower normalization band,
much smaller than expected in \citet[][see also Fig.~13]{Hoff96}. The
contributions to the production of $^{45}$Sc and $^{49}$Ti are not
important either, which are suggested to be produced in the earlier
ejecta \citep{Prue05, Froh06}.

Figure~17 compares the results with (\textit{top panels}) and without
(\textit{bottom panels}) neutrino reactions (eqs.~[1]-[6]) for the $M =
1.4\, M_\odot$ (\textit{left panels}) and $2.0\, M_\odot$ (\textit{right
panels}) models. As can be seen, the production of almost all $^{92}$Mo
and some portions of $^{74}$Se, $^{78}$Kr, and $^{84}$Sr are not due to
the effect of neutrino reactions. This can be also seen in Figure~18,
which compares the results from only neutron-rich (\textit{top panels})
or proton-rich (\textit{bottom panels}) winds for the $M = 1.4\,
M_\odot$ (\textit{left panels}) and $2.0\, M_\odot$ (\textit{right
panels}) models. $^{92}$Mo is highly overproduced in the mass average of
neutron-rich winds, which is, on the contrary, somewhat underproduced in
the proton-rich winds compared to other light \textit{p}-nuclei (e.g.,
$^{74}$Se, $^{78}$Kr, $^{84}$Sr, $^{94}$Mo, and $^{96, 98}$Ru). Hence,
the bulk of $^{92}$Mo originates from the slightly neutron-rich winds
($Y_{ef} \approx 0.46-0.50$), while other \textit{p}-nuclei from mainly
proton-rich winds.

When we look at the top panels of Figure~17 again, the
\textit{p}-nuclei up to $^{92}$Mo and $^{108}$Cd for $M=1.4\, M_\odot$
and $2.0\, M_\odot$ models, respectively, fall within the
normalization band, which are regarded to be the dominant species
produced by each event. It is interesting to note that the
contributions to $^{94}$Mo, $^{96, 98}$Ru, and $^{102}$Pd for $M =
1.4\, M_\odot$ and $^{112, 114, 115}$Sn, and $^{113}$In for $M = 2.0\,
M_\odot$ are marginal. The ejected masses by winds during the first
20~s are $3.0 \times 10^{-3}\, M_\odot$ and $1.2 \times 10^{-3}\,
M_\odot$ for the current $M = 1.4\, M_\odot$ and $2.0\, M_\odot$
models, respectively. Given that the progenitor mass for each case to
be, e.g., $15\, M_\odot$ and $30\, M_\odot$, respectively, the
overproduction factor is expressed as $\sim 10^{-4}\,
(X_\mathrm{ej}/X_\odot)$. The requisite overproduction factor for the
nucleosynthetic event to be the major source in the solar system is
$\sim 10$ \citep{Woos94}, assuming that \textit{all} the core-collapse
supernovae produce the same amount of the isotope. This number
increases (e.g., $\sim 100$) if a limited fraction of supernovae
(e.g., $25-30 M_\odot$) contribute to these light \textit{p}-nuclei
production \citep[see][for the \textit{r}-process elements]{Ishi99,
Wana06b}. This is conceivable since the current nucleosynthesis
results are highly dependent on the physical conditions such as
entropy or $Y_e$. The overproduction factors of $\sim 10-100$
(Fig.~17, \textit{top panels} or Fig.~15) for the current models imply
that the neutrino-driven winds can be potentially the major
astrophysical site of these light \textit{p}-nuclei.

\section{Summary and Conclusions}

In this study, the neutrino-induced \textit{rp}-process in
neutrino-driven winds has been investigated. The thermodynamic histories
of winds were obtained from the semi-analytic models of the
neutrino-driven winds that had been developed for the \textit{r}-process
study in previous works. The subsonic wind solutions with the nearly
maximum mass ejection rates were taken in this study, for various
neutrino luminosities ($40-0.5 \times 10^{51}\, \mathrm{ergs\ s}^{-1}$)
with the proto-neutron star masses of $1.4\, M_\odot$ and $2.0\,
M_\odot$. The latter ($2.0\, M_\odot$) was regarded as the test case of
higher entropy winds (about a factor of two), which might be expected in
more realistic hydrodynamic simulations of core-collapse supernovae. The
nucleosynthesis calculations were performed for the wide range of the
initial electron fractions ($0.45 \le Y_{e0} \le 0.65$) including rather
proton-rich compositions, motivated by recent hydrodynamic results. The
main results of this study can be summarized as follows:

\noindent 1. In the proton-rich winds, the proton capture can proceed
beyond the iron-group nuclei, by bypassing the known $\beta^+$-waiting
point nuclei (e.g., $^{64}$Ge) via $(n,\, p)$ reactions, as also
suggested by recent works \citep{Froh06, Froh06b, Prue06, Wana06}. These
neutrons are continuously supplied from the anti-electron neutrino
capture on abundant free protons.

\noindent 2. The neutrino-induced \textit{rp}-process leads to the
production of some light \textit{p}-nuclei, even with moderate entropy ($\sim
50\, N_Ak$) that can be found in the early winds ($t_\mathrm{pb} \sim
1\, \mathrm{s}$) from the neutron star with $1.4\, M_\odot$. The
production of \textit{p}-nuclei is, however, highly dependent on the entropy in
the wind, which affects the proton-to-seed abundance ratio $Y_p/Y_h$ at
the beginning of the \textit{rp}-process. In the high entropy winds
($\sim 100\, N_Ak$) with $M = 2.0\, M_\odot$, the nuclear flow proceeds
to $Z \approx 50$, resulting in the production of \textit{p}-nuclei up to $A
\sim 130$.

\noindent 3. The production of \textit{p}-nuclei is also highly
sensitive to the proton richness of the wind material. In the slightly
proton-rich winds ($Y_e \sim 0.50-0.52$), only the lightest
\textit{p}-nuclei ($^{74}$Se, $^{78}$Kr, and $^{84}$Sr) are
produced. The higher \textit{p}-nuclei up to $A \sim 130$ are
synthesized with interesting amounts as $Y_e$ rises to $\sim 0.56$.

\noindent 4. The termination of the \textit{rp}-process is due to the
increasing Coulomb barrier for the heaviest nuclei in the nuclear flow,
\textit{not} due to the exhaustion of free protons. The decrease of the
temperature below $\sim 2 \times 10^9\, \mathrm{K}$ owing to the
expansion of the ejecta is also a cause of the termination. The supply
of neutrons by the neutrino capture on protons is sufficient throughout
the \textit{rp}-process phase. Hence the dilution of the neutrino flux
by the expansion of matter is not the principal reason for the
termination of the \textit{rp}-process in the current wind models.

\noindent 5. Few \textit{p}-nuclei beyond $A \sim 130$ are produced by
the neutrino-induced \textit{rp}-process, owing to the increasing
Coulomb barrier for heavier nuclei. The high entropy ($\sim 200\, N_Ak$)
or highly proton-rich ($Y_e \sim 0.6$) wind results in driving the
nuclear abundances to the $\beta$-stability or even the neutron-rich
region, when the temperature drops below $T_9 \sim 2$. In this case, a
variety of neutron-rich isotopes, which are generally assign to
\textit{s}-nuclei or \textit{r}-nuclei, are synthesized even in the
proton-rich compositions.

\noindent 6. $^{92}$Mo is produced with interesting amounts in the
slightly neutron-rich winds ($Y_e \sim 0.46-0.49$), much more than in
the proton-rich winds. This is due to the proton capture reactions at $T
\sim 4-3 \times 10^9\, \mathrm{K}$, which carry a portion of the nuclear
abundances from the abundant $^{90}$Zr to $^{92}$Mo. The amount of the
produced $^{92}$Mo is, however, highly dependent on the thermodynamic
history of the wind.

\noindent 7. The neutrino-driven winds can be the origin of some light
\textit{p}-nuclei, at least of $^{74}$Se, $^{78}$Kr, $^{84}$Sr, and
$^{92}$Mo, and likely of $^{94}$Mo, $^{96, 98}$Ru, $^{102}$Pd, and
$^{106, 108}$Cd, supplemented by the production of $^{92}$Mo in the
slightly neutron-rich compositions. In particular, this can be the
unique astrophysical site responsible for the production of the
proton-rich Mo and Ru isotopes. The heavier \textit{p}-nuclei ($A >
110$) may have another origin, most likely the O/Ne layers in
core-collapse supernovae \citep{Pran90, Raye95}. Note that the
\textit{p}-nuclei synthesized in neutrino-driven winds are regarded as
\textit{primary}. This is contrast to those produced in the O/Ne layers,
which need the \textit{s}-process seed abundances and thus regarded as
\textit{secondary}.

Note that the neutrino-induced \textit{rp}-process in the current
study may not contribute to the production of any \textit{elements},
since the fraction of \textit{p}-process isotopes for a given element
is generally an order of one percent. Therefore, it would be
challenging to confirm the occurrence of the neutrino-induced
\textit{rp}-process by spectroscopic studies of extremely metal-poor
stars, which are expected to preserve the nucleosynthetic signature
from a single (or a few) supernova \citep[e.g.,][]{Ishi99,
Wana06b}. Analysis of the isotope anomalies in some elements found in
primitive meteorites \citep[e.g.,][]{Arno03} or measurements of
Galactic cosmic-rays might provide us some clues.

At the end of this investigation, it should be cautioned that the
nucleosynthesis calculations in the current study are based on a
schematic formulation of neutrino-driven winds, which must be in fact
closely related to the unknown mechanism of core-collapse
supernovae. In addition, the estimation of the \textit{p}-nuclei
production in each event (i.e., the core-collapse supernova that
leaves the proto-neutron star with $1.4\, M_\odot$ or $2.0\, M_\odot$)
was based on the $Y_e$ distribution during the \textit{bubble} (not
wind) phase, obtained from one specific hydrodynamic simulation
\citep{Bura06}. Therefore, the current results should \textit{not} be
regarded as the quantitative predictions that is to be used for the
Galactic chemical evolution study.

Nevertheless, the current study provides us some notable
implications. First, it is likely that the neutrino-induced
\textit{rp}-process takes place in \textit{all} core-collapse
supernovae to some extent, even with a moderate entropy. In this
sense, other astrophysical sites than standard core-collapse events,
e.g., collapsar jets or disk winds formed around a black hole, which
are associated to gamma-ray bursts, can be the astrophysical site for
the neutrino-induced \textit{rp}-process. Second, details of the
nucleosynthesis process are presented for selected winds, which will
be useful for the future experimental studies of proton-rich nuclei
far from the $\beta$-stability. In particular, core-collapse
supernovae no doubt eject the nucleosynthetic products that contribute
to the Galactic chemical evolution of heavy nuclei. This is in
contrast to X-ray bursts, which is unlikely to contribute the Galactic
chemical evolution. Therefore the experimental estimations can be
easily tested by comparing the nucleosynthesis calculations to, e.g.,
the solar compositions. Third, the systematic calculations for the
wide (but reasonable) ranges of the neutrino luminosities and the
electron fractions enable us to make a meaningful comparison with the
solar abundances. In particular, a notable aspect is that $^{92}$Mo
has another origin (i.e., neutron-rich winds) while other light
\textit{p}-nuclei are mainly synthesized by the \textit{rp}-process
(i.e., in proton-rich winds) even in a \textit{single} event. This
will serve unique constraints to the fluid dynamics of the early
supernova ejecta.

\acknowledgements

This work was supported in part by a Grant-in-Aid for Japan-France
Integrated Action Program (SAKURA) from the Japan Society for the
Promotion of Science, and Scientific Research (17740108) from the
Ministry of Education, Culture, Sports, Science, and Technology of
Japan.

%\clearpage

\begin{figure}
\epsscale{1.}
%\plotone{lnuf.ps}
\plotone{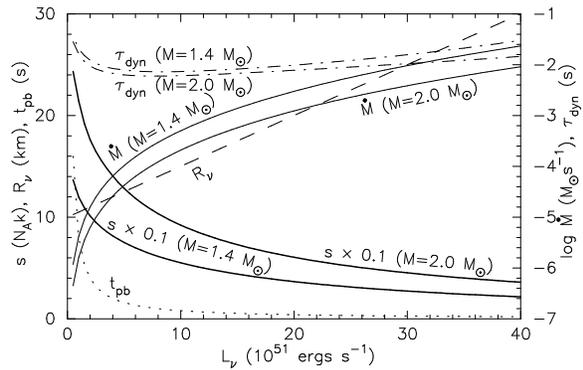}

\caption{Model parameters ($R_\nu, t_\mathrm{pb}, \dot M$) of the
  current neutrino-driven winds are shown as functions of $L_\nu$. Also
  denoted are the obtained entropies ($s$) and dynamic timescales
  ($\tau_\mathrm{dyn}$) for $1.4\, M_\odot$ and $2.0\, M_\odot$ cases.  }

\end{figure}

\clearpage

\begin{figure}
\epsscale{1.8}
%\plotone{tj14.ps}
\plotone{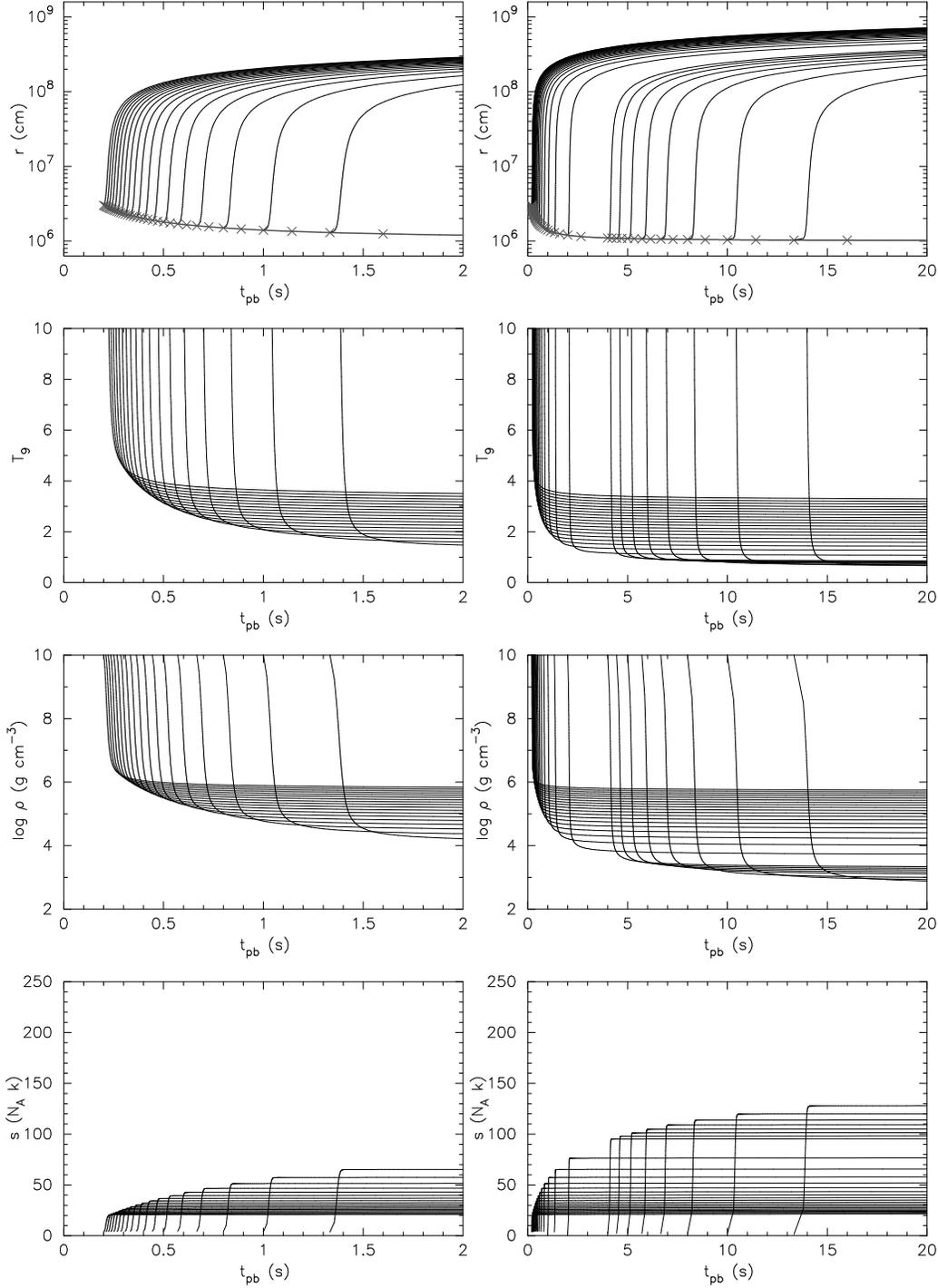}

\caption{Time variations of radius, temperature, density, and entropy
for selected (odd-numbered) wind trajectories with the $M = 1.4\,
M_\odot$ model ($t_\mathrm{pb} \le 2\, \mathrm{s}$ and $t_\mathrm{pb}
\le 20\, \mathrm{s}$ for \textit{left} and \textit{right panels},
respectively). The evolution of the neutrino sphere is denoted by the
grey lines along with the staring point for each wind (crosses) in top
panels.}

\end{figure}

\clearpage

\begin{figure}
\epsscale{1.8}
%\plotone{tj20.ps}
\plotone{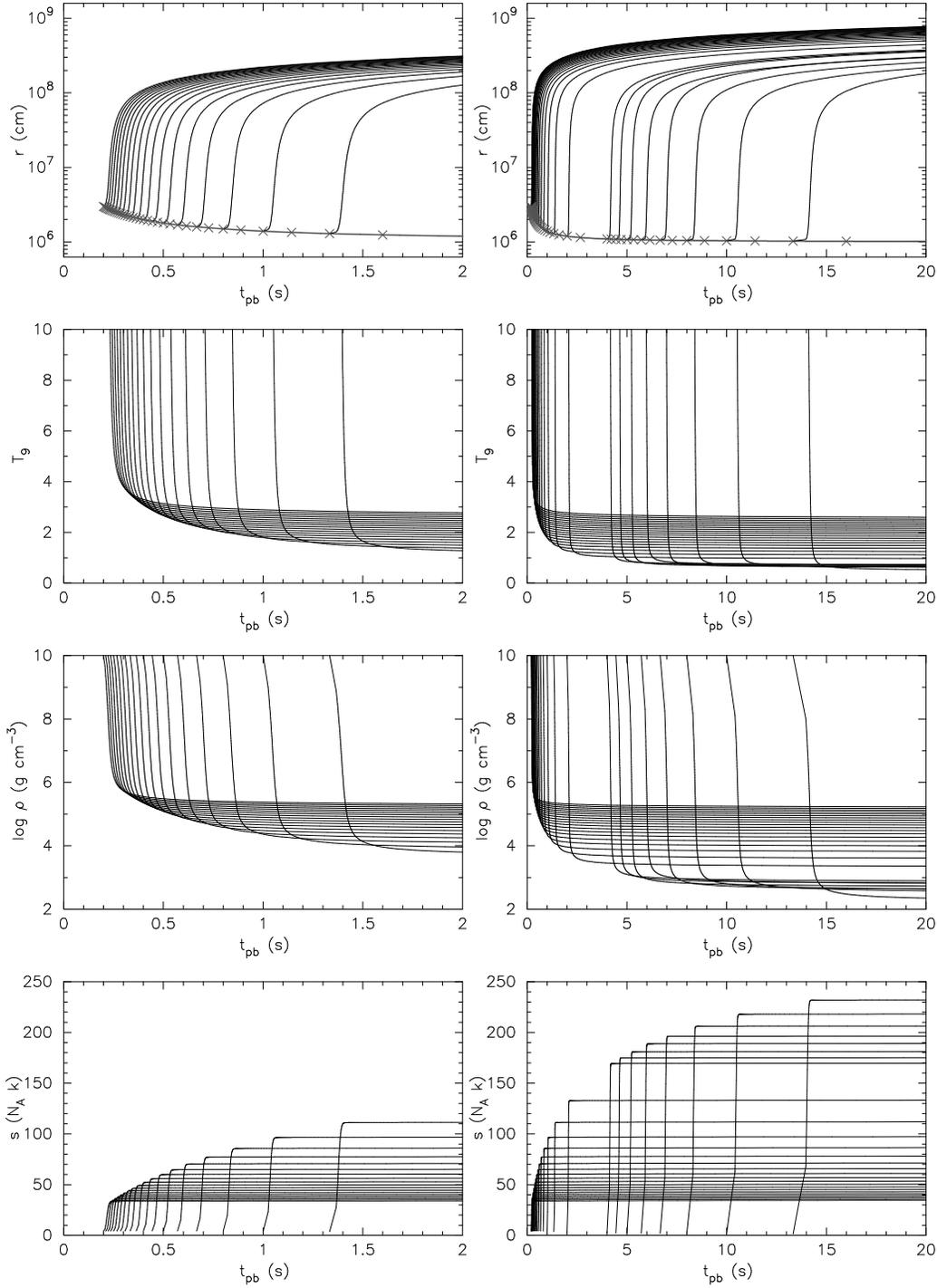}

\caption{Same as Figure~2, but for the $M = 2.0\, M_\odot$ model.}

\end{figure}

\clearpage

\begin{figure}
\epsscale{1.0}
%\plotone{yet.ps}
\plotone{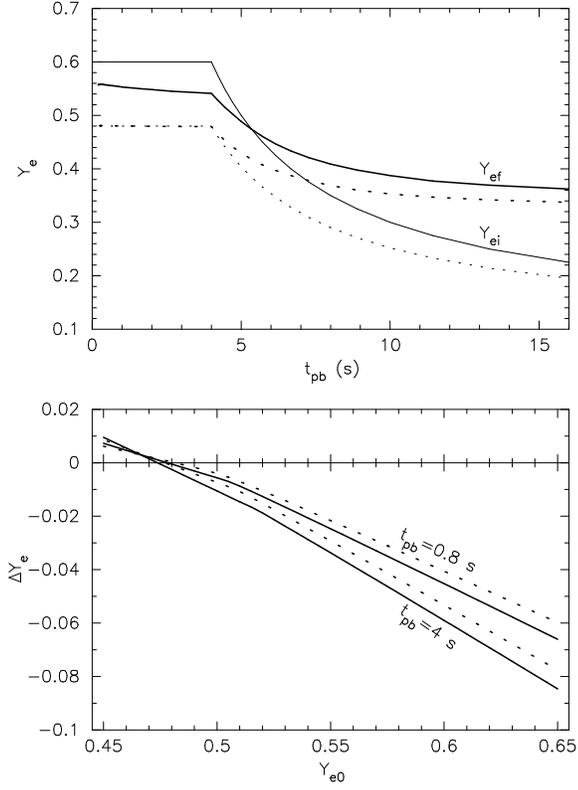}

\caption{\textit{Top}: $Y_e$ at $T_9 = 9$ ($Y_{ei}$, \textit{thin
 lines}) and at $T_9 = 3$ ($Y_{ef}$, \textit{thick lines}) as functions
 of $t_\mathrm{pb}$. The solid and dotted lines denote the results for
 $M = 2.0\, M_\odot$ with $Y_{e0} = 0.60$ and $M = 1.4\, M_\odot$ with
 $Y_{e0} = 0.48$, respectively. \textit{Bottom}: Differences of $Y_e$
 ($\mathit{\Delta} Y_e \equiv Y_{ei} - Y_{ef}$) at $t_\mathrm{pb} =
 0.8\, \mathrm{s}$ and $4.0\, \mathrm{s}$ as functions of $Y_{e0}$, for
 $M = 1.4\, M_\odot$ (\textit{dotted lines}) and $2.0\, M_\odot$
 (\textit{solid lines}).
}

\end{figure}

%\clearpage

\begin{figure}
\epsscale{1.}
%\plotone{tev0.ps}
\plotone{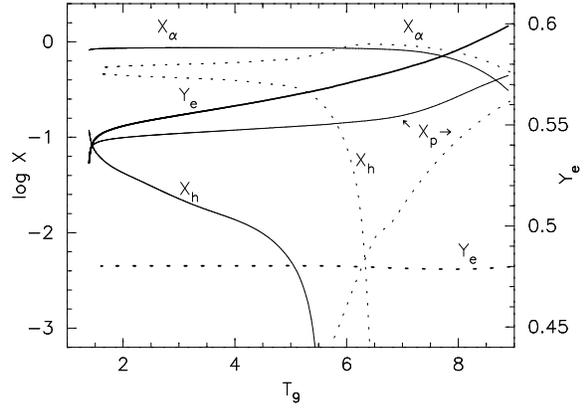}

\caption{Variations of $Y_e$ (\textit{thick lines}) and the mass
fractions (\textit{thin lines}) of neutrons ($X_n$), protons ($X_p$),
$\alpha$ particles ($X_\alpha$), and heavy nuclei ($X_h$) as functions
of $T_9$. The solid and dotted lines denote the results for the wind
trajectories W201060 and W141048 (see the text and footnote~1),
respectively.}

\end{figure}

\clearpage

\begin{figure}
\epsscale{1.8}
%\plotone{nz.ps}
\plotone{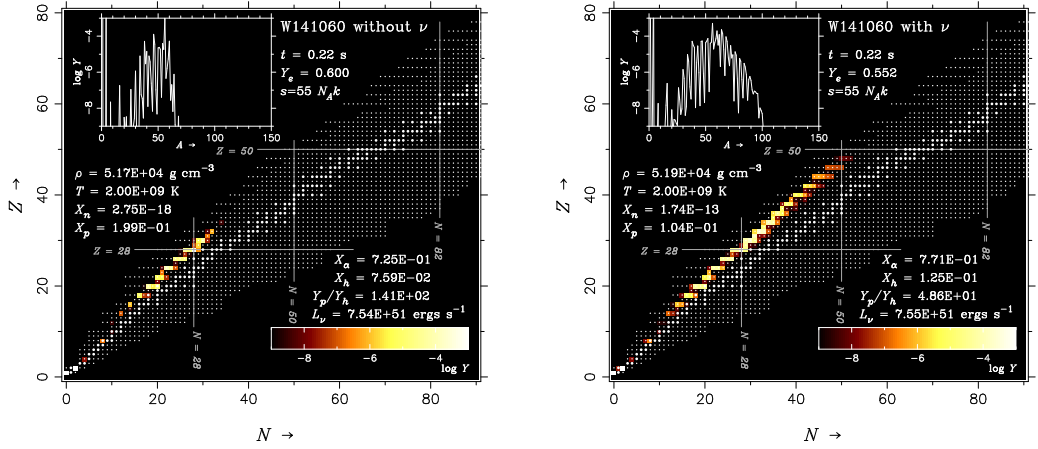}

\caption{Snapshots of the nucleosynthesis calculations at $t = 0.22\,
  \mathrm{s}$ for the wind trajectory of W141060, in which the
  neutrino-induced reactions (eqs.~[1]-[6]) are turned off (\textit{left
  panel}) and on (\textit{right panel}). The temperature decreases to
  $T_9 \approx 2$ at this time. The abundances are color coded in the
  nuclide chart. The nuclei included in the reaction network are denoted
  by dots, with the stable isotopes represented by large dots. The
  abundance curve as a function of mass number is shown in the upper
  left for each panel.}

\end{figure}

\clearpage

\begin{figure}
\epsscale{1.8}
%\plotone{nz2.ps}
\plotone{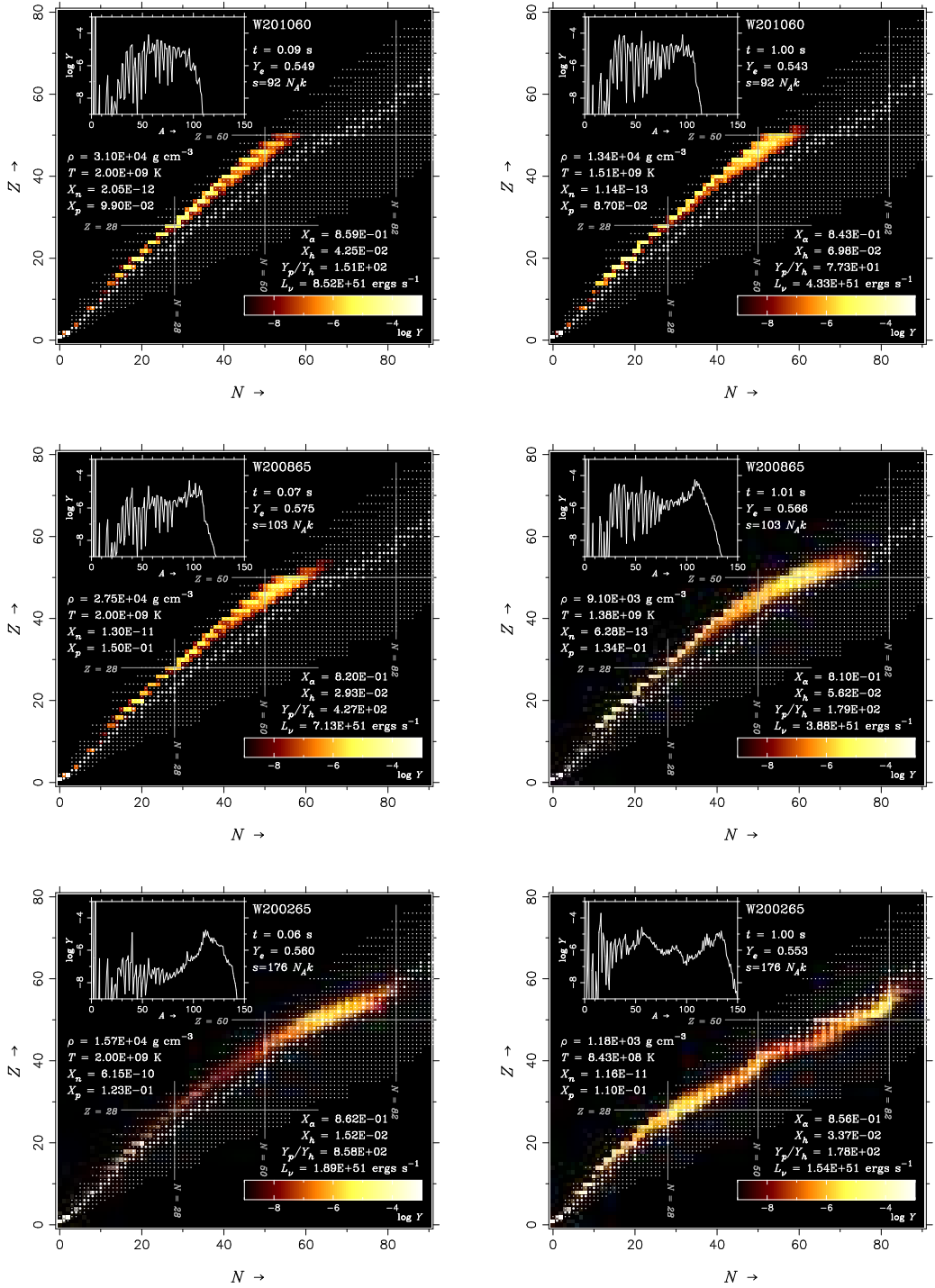}

\caption{Same as Figure~6, but for the wind trajectories of W201060
(\textit{top}), W200865 (\textit{middle}), and W200265
(\textit{bottom}). Left and right panels are the snapshots at $T_9
\approx 2.0$ and $t_\mathrm{pb} \approx 1.0\, \mathrm{s}$, respectively. 
Neutrino-induced reactions (eqs.~[1]-[6]) are included for all cases.}

\end{figure}

\clearpage

\begin{figure}
\epsscale{1.}
%\plotone{flow.ps}
\plotone{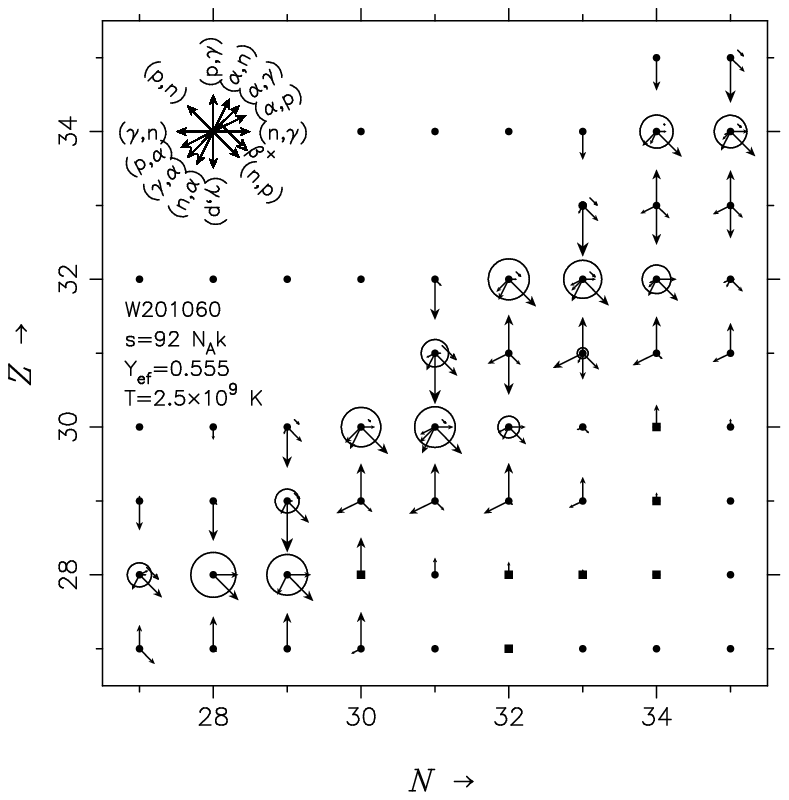}

\caption{Nuclear flows defined by equation~(9) (\textit{arrows}) and the
 abundances (\textit{circles}) in logarithmic scale for the wind
 trajectory of W201060 at $T_9 = 2.5$, for those greater than $10^{-5}$.
 The flows by $\beta^+$-decays are denoted by off-centered arrows. The
 nuclei included in the reaction network are denoted by dots, with the
 stable isotopes represented by squares.}

\end{figure}

%\clearpage

\begin{figure}
\epsscale{1.}
%\plotone{tev.ps}
\plotone{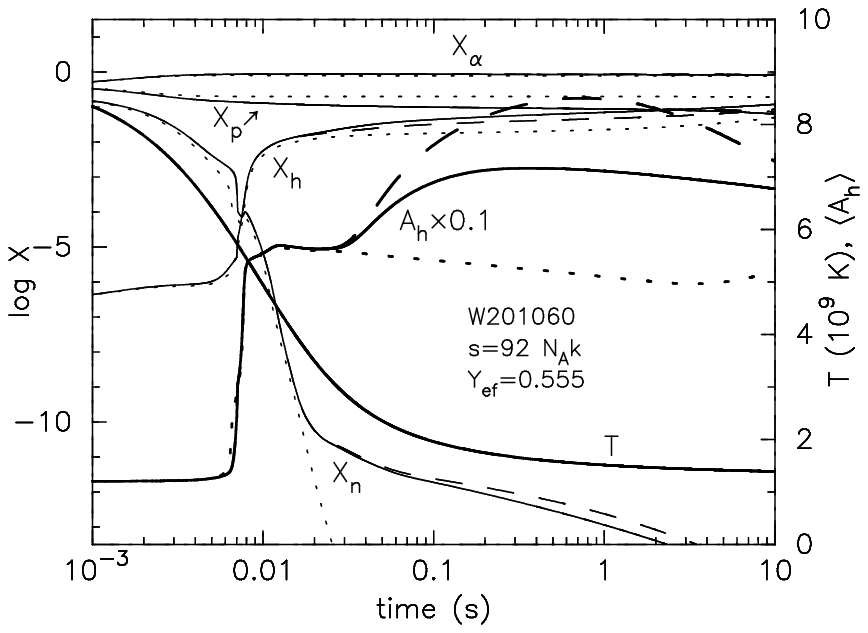}

\caption{Time variations of the mass fractions (\textit{thin lines}) of
neutrons $X_n$, protons $X_p$, $\alpha$ particles $X_\alpha$, and heavy
nuclei $X_h$ for the wind trajectory W201060. The solid, dashed, and
dotted lines denote the results with all neutrino-induced reactions
(eqs.~[1]-[6]), neutrino capture on free nucleons only (eqs.~[1] and
[2]), and no neutrino-induced reactions, respectively. Also shown are
the average mass number of heavy nuclei $A_h$ and the temperature
(\textit{thick lines}).  }

\end{figure}

%\clearpage

\begin{figure}
\epsscale{1.}
%\plotone{tau.ps}
\plotone{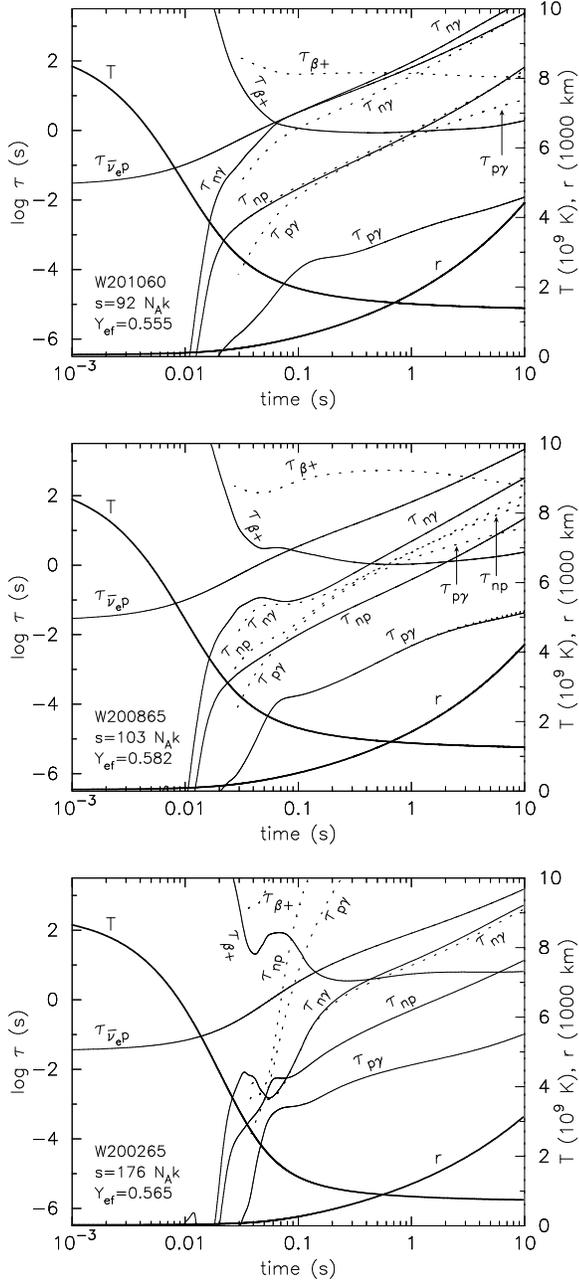}

\caption{Time variations of the mean lifetimes for $\beta^+$-decay, $(n,
\gamma)$, $(p, \gamma)$, and $(n, p)$ reactions, and neutrino capture on
protons for the wind trajectories of W201060 (\textit{top}), W200865
(\textit{middle}), and W200265 (\textit{bottom}), defined by
equations~(12)-(16). The solid and dotted lines denote the lifetimes for
$Z \ge 2$ and $Z \ge 48$ nuclei, respectively. For the latter, the
values for only $T_9 < 3$ are shown, in which these nuclei are
synthesized by the \textit{rp}-process. Also shown are the temperature
and the distance from the center of the neutron star (\textit{thick
lines}).}

\end{figure}

%\clearpage

\begin{figure}
\epsscale{1.}
%\plotone{flow2.ps}
\plotone{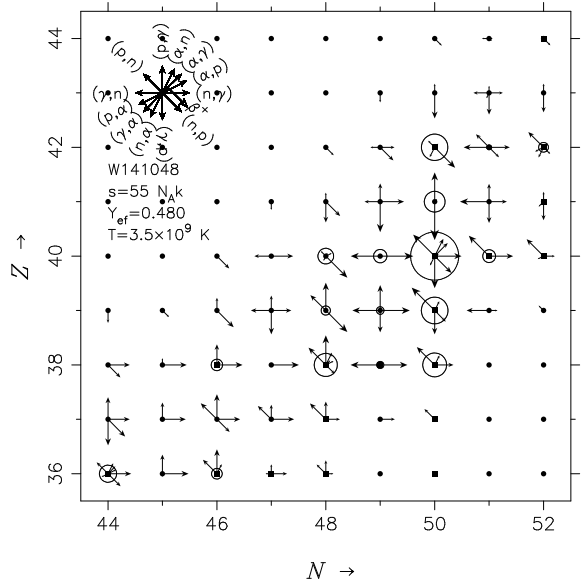}

\caption{Same as Figure~8, but for the wind trajectory of W141048.  }

\end{figure}

%\clearpage

\begin{figure}
\epsscale{1.}
%\plotone{tev2.ps}
\plotone{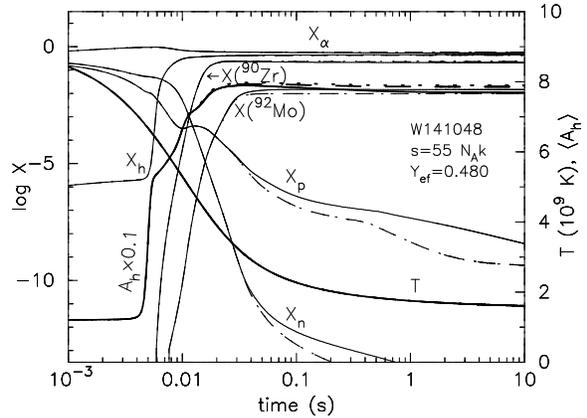}

\caption{Same as Figure~9, but for the wind trajectory of W141048. The
 mass fractions of $^{90}$Zr and $^{92}$Mo are also plotted.}

\end{figure}

%\clearpage

\begin{figure}
\epsscale{2.}
%\plotone{oproye.ps}
\plotone{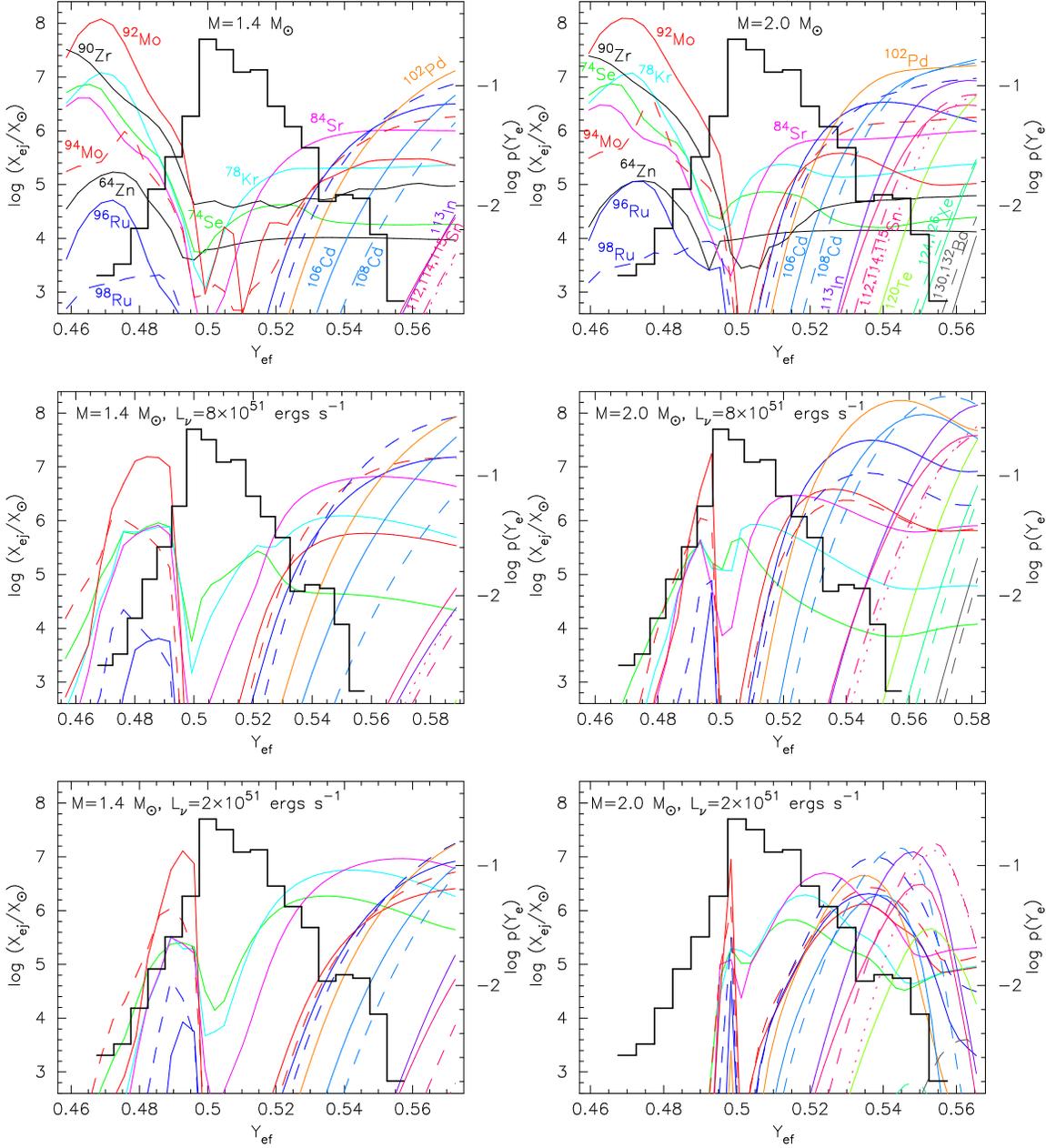}

\caption{Mass fractions $X_\mathrm{ej}$ of \textit{p}-nuclei with
respect to their solar values \citep[$X_\odot$,][]{Ande89} as functions
of $Y_{ef}$ for $M = 1.4\, M_\odot$ (\textit{left}) and $2.0\, M_\odot$
(\textit{right}) models. Top panels show the mass-averaged results (see
the text), where $Y_{ef}$ is taken to be that at $t_\mathrm{pb} = 4\,
\mathrm{s}$ ($L_\nu = 2 \times 10^{51}\, \mathrm{ergs\ s}^{-1}$). The
mass fractions of $^{64}$Zn and $^{90}$Zr are also plotted for
comparison purposes. Middle and bottom panels show the results for the
winds with $L_\nu = 8 \times 10^{51}\, \mathrm{ergs\ s}^{-1}$ and $2
\times 10^{51}\, \mathrm{ergs\ s}^{-1}$, respectively. The histogram is
the asymptotic $Y_e$ distribution $p(Y_e)$ of the neutrino-processed
ejecta during the first 468~ms after core bounce for a $15\, M_\odot$
progenitor star taken from \cite{Bura06}.}

\end{figure}

\clearpage

\begin{figure}
\epsscale{2.}
%\plotone{oprolnu.ps}
\plotone{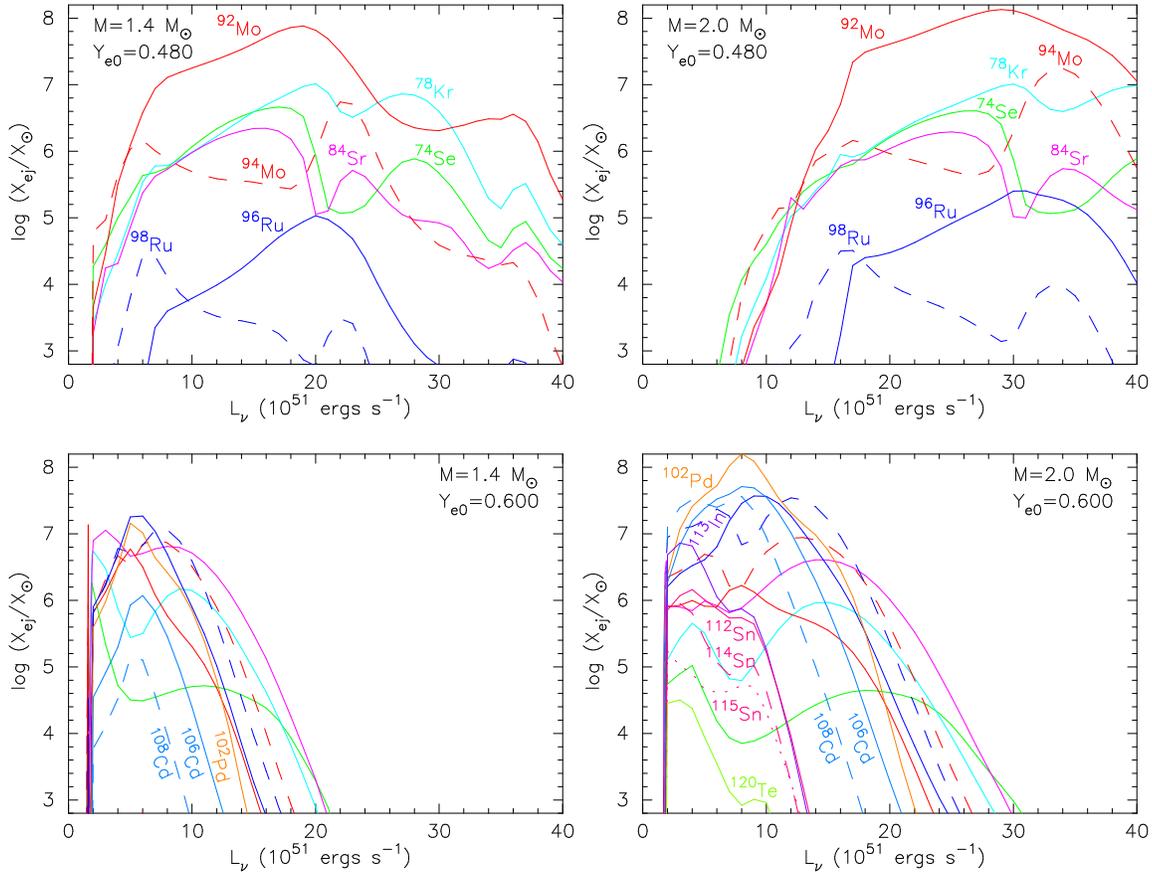}

\caption{Mass fractions of \textit{p}-nuclei with respect to their solar
values \citep{Ande89} as functions of $L_\nu$, for the selected cases
($M = 1.4\, M_\odot$; \textit{left}, $M = 2.0\, M_\odot$;
\textit{right}, $Y_{e0} = 0.48$; \textit{top}, $Y_{e0} = 0.60$;
\textit{bottom}).  }

\end{figure}

\clearpage

\begin{figure}
\epsscale{2.}
%\plotone{oprone.ps}
\plotone{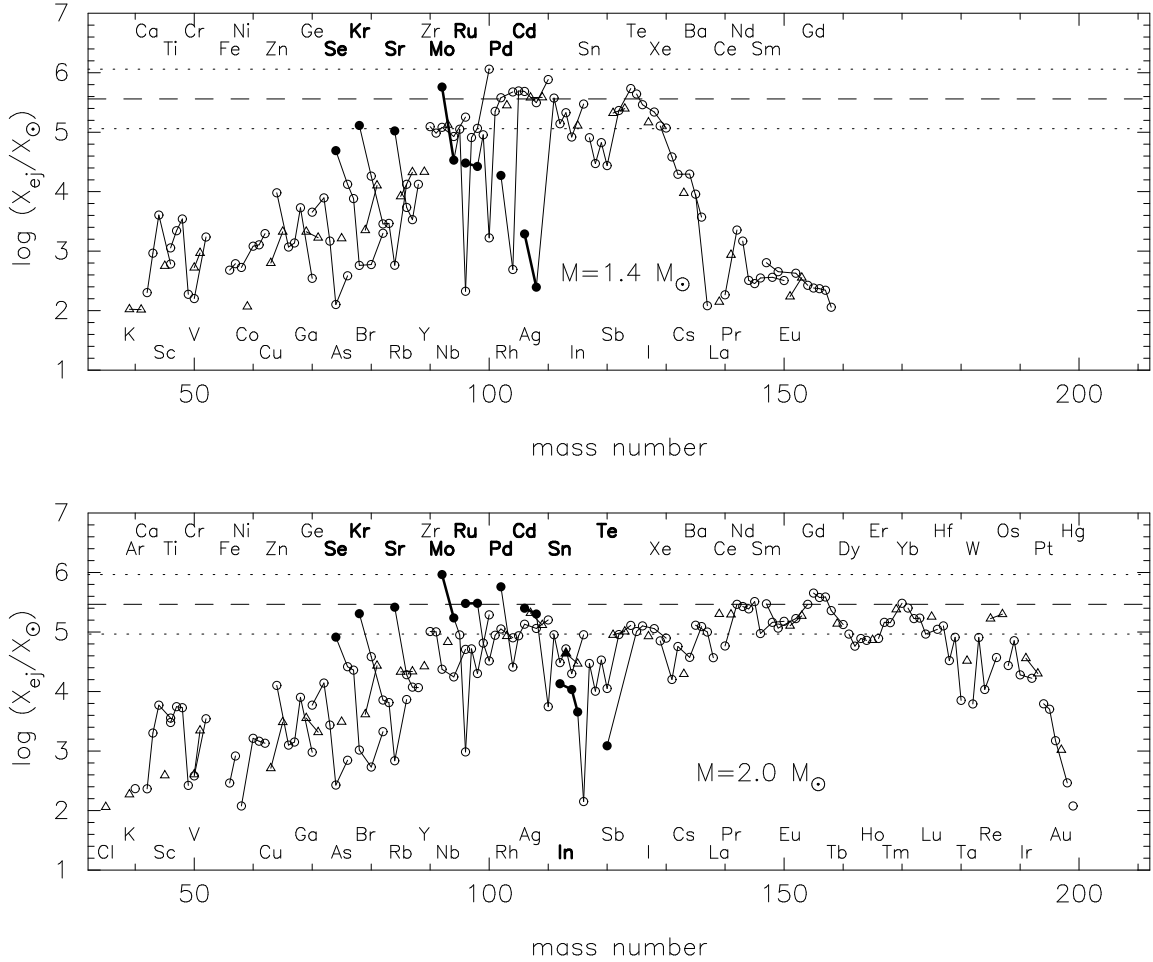}

\caption{Mass-$Y_e$-averaged abundances with respect to their solar
values \citep{Ande89} for $M = 1.4\, M_\odot$ (\textit{top}) and $2.0\,
M_\odot$ (\textit{bottom}) models as functions of mass number (see the
text). The abundances smaller than $X_\mathrm{ej}/X_\odot < 100$ is
omitted here. The isotopes (after decay) are denoted by open circles
(even-$Z$) and triangles (odd-$Z$). The \textit{p}-nuclei are denoted
with filled symbols. The solid lines connect isotopes of a given
element.  }

\end{figure}

\clearpage

\begin{figure}
\epsscale{2.}
%\plotone{oprone.ps}
\plotone{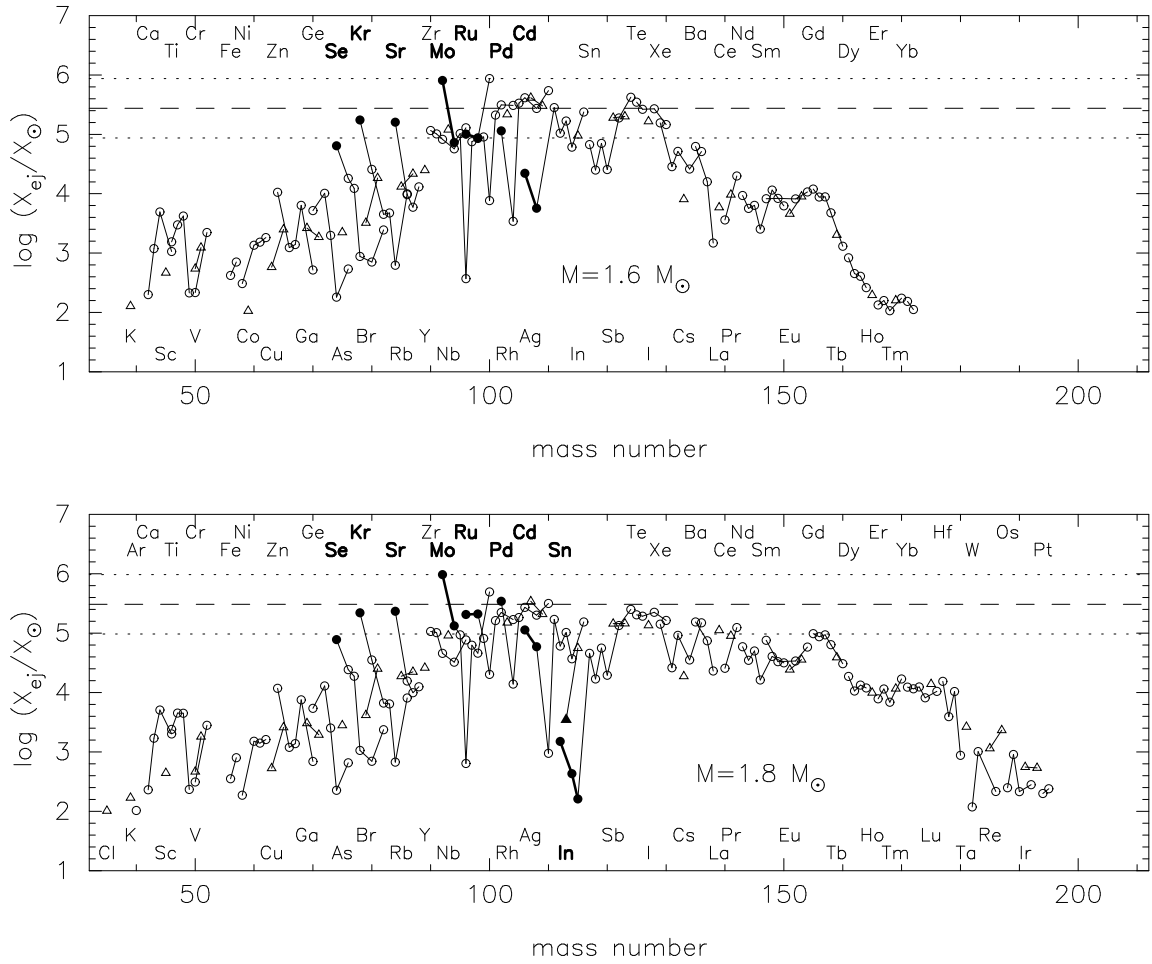}

\caption{Same as Figure~15, but for $M = 1.6\, M_\odot$ (\textit{top}) and $1.8\,
M_\odot$ (\textit{bottom}) models.  }

\end{figure}

\clearpage

\begin{figure}
\epsscale{2.}
%\plotone{oprone2.ps}
\plotone{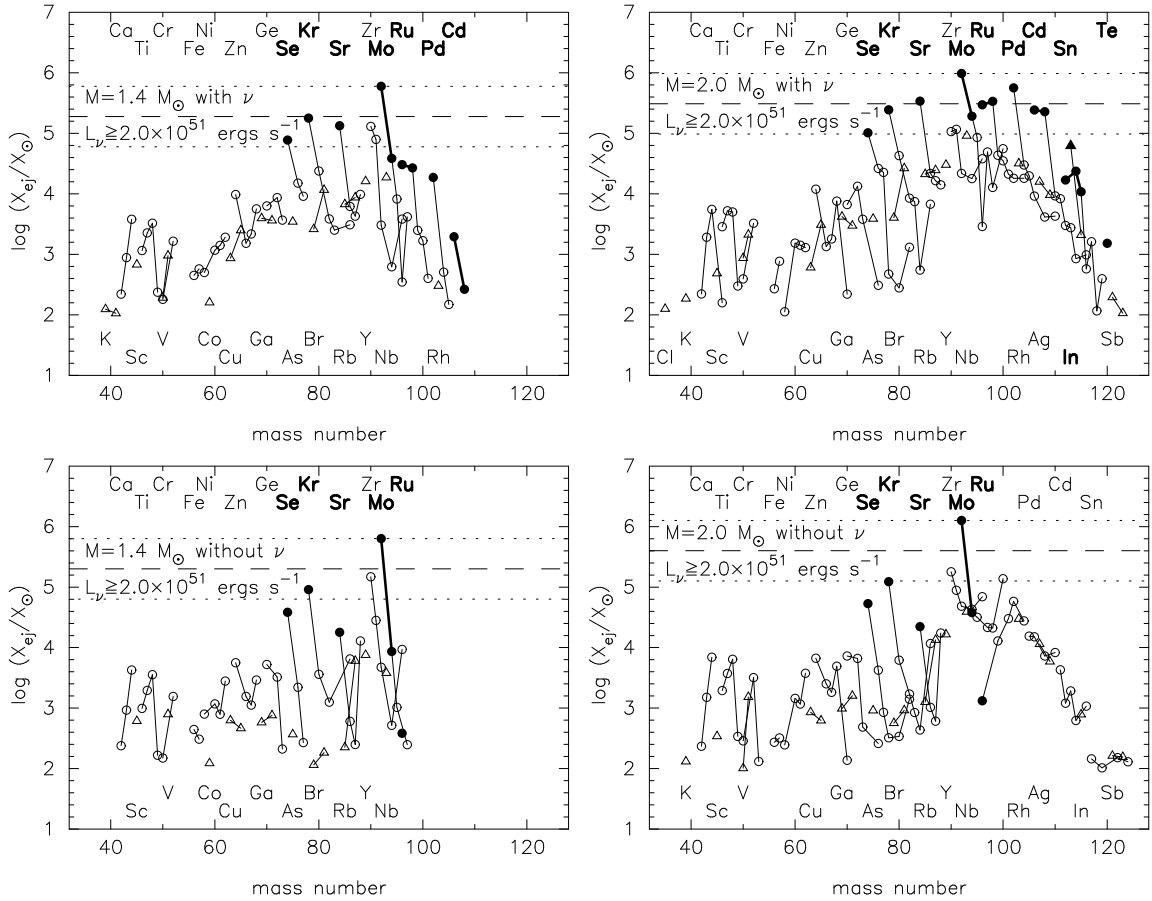}

\caption{Same as Figure~15, but for $L_\nu \ge 2 \times 10^{51}\,
\mathrm{ergs\ s}^{-1}$ ($t_\mathrm{pb} \le 4\, \mathrm{s}$). Top and
bottom panels show the results with and without neutrino-induced
reactions (eqs.~[1]-[6]), respectively.}

\end{figure}

\clearpage

\begin{figure}
%\epsscale{2.}
%\plotone{oprone3.ps}
\plotone{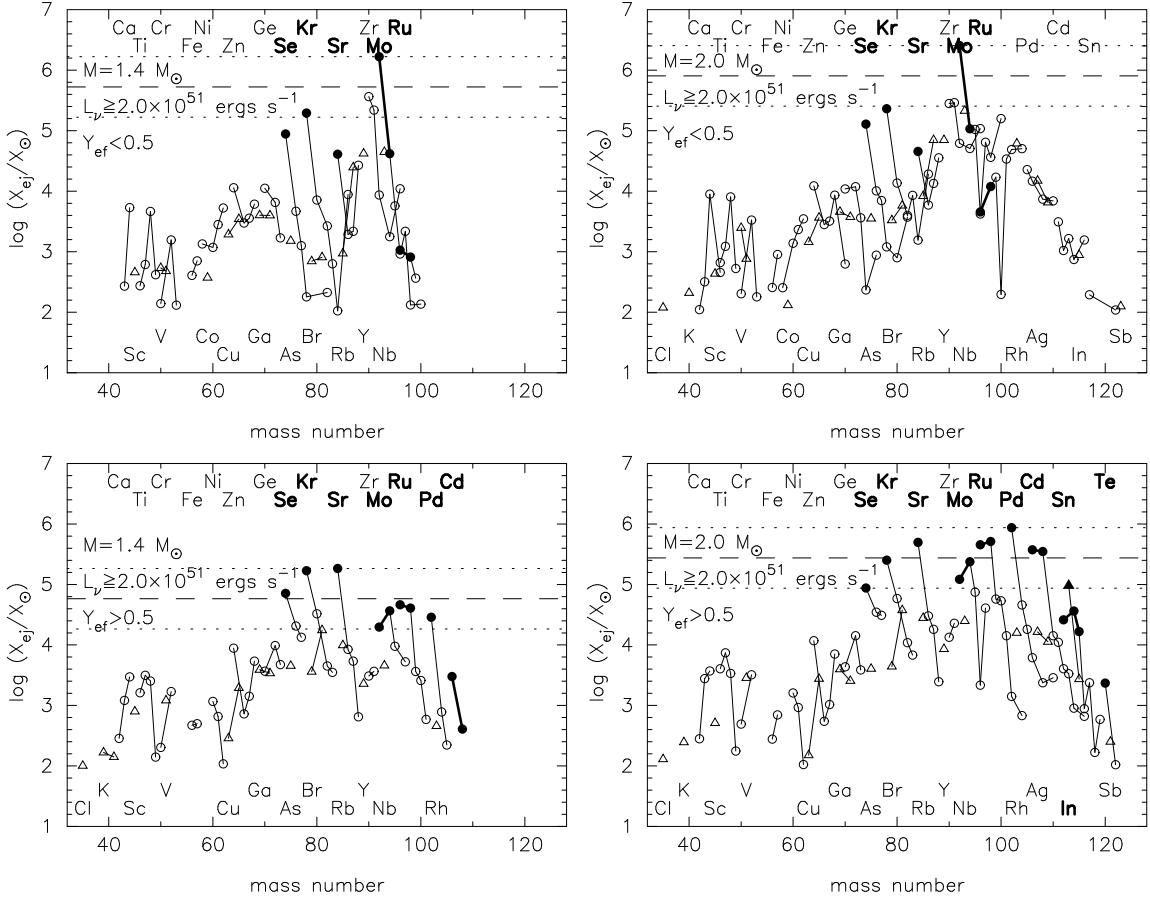}

\caption{Same as Figure~15, but for $L_\nu \ge 2 \times 10^{51}\,
\mathrm{ergs\ s}^{-1}$ ($t_\mathrm{pb} \le 4\, \mathrm{s}$). Top and
bottom panels show the results for the neutron-rich ($Y_{ef} < 0.5$) and
proton-rich ($Y_{ef} > 0.5$) winds, respectively.}

\end{figure}

\end{document}